\documentclass[12pt,3p]{elsarticle}
\pdfoutput=1
\biboptions{numbers,sort&compress}

\usepackage{graphicx,float}
\usepackage{color, graphics}
\usepackage{amsmath,amssymb,amsfonts}
\usepackage{verbatim}   
\usepackage{subfigure}  
\usepackage{acronym,afterpage}
\usepackage{amsmath}
\usepackage{hyperref}
\usepackage{mathrsfs}
\usepackage[labelsep=endash]{caption}
\usepackage{multirow}
 \usepackage{cancel}
 \usepackage{url}
  \usepackage{slashed}
  \usepackage{ tipa }
 \usepackage{hyperref}
 \usepackage{nicefrac}
\usepackage{pifont}

\newcommand{\xmark}{\ding{55}}

\def\beq{\begin{equation}\begin{aligned}}
\def\eeq{\end{aligned}\end{equation}}

\journal{Nuclear Physics B}

\begin{document}

\begin{frontmatter}

\title{Axial Vector $Z'$ and Anomaly Cancellation}

\author[label1,label2,label3]{Ahmed Ismail}
\address[label1]{Department of Physics, University of Illinois at Chicago, Chicago, IL 60607, USA}
\address[label2]{High Energy Physics Division, Argonne National Laboratory, Argonne, IL 604339, USA}
\address[label3]{Department of Physics and Astronomy, University of Pittsburgh, Pittsburgh, PA 15260, USA}

\author[label1]{Wai-Yee Keung}

\author[label1]{Kuo-Hsing Tsao \corref{cor1}}
\cortext[cor1]{Corresponding author}
\ead{ktsao2@uic.edu}

\author[label1]{James Unwin }

\tnotetext[t1]{PITT-PACC-1607}

\begin{abstract}
Whilst the prospect of new $Z'$ gauge bosons with only axial couplings to the Standard Model (SM) fermions is widely discussed, examples of anomaly-free renormalisable models are lacking in the literature. We look to remedy this by constructing several motivated examples. Specifically, we consider axial vectors which couple universally to all SM fermions, as well as those which are generation-specific, leptophilic, and leptophobic. Anomaly cancellation typically requires the presence of new coloured and charged chiral fermions, and we argue that in a large class of models masses of these new states are expected to be comparable to that of the axial vector. Finally, an axial vector mediator could provide a portal between SM and hidden sector states, and we also consider the possibility that the axial vector couples to dark matter. If the dark matter relic density is set due to freeze-out via the axial vector, this strongly constrains the parameter space. 

\end{abstract}

\end{frontmatter}
\renewcommand*{\today}{September 7, 2016}


\vspace{3mm}
\section{Introduction}
\label{Sec1}

Couplings between chiral fermions $f_L$, $f_R$ and a vector boson $Z'$ associated to a U(1) gauge symmetry are of the form 
\beq
\overline{f}\slashed{D}f
=&\overline{f}\gamma_\mu\left(\partial^\mu
-ig(q_{f_L}+q_{f_R})\frac{1}{2}Z'{}^\mu-i g(q_{f_L}-q_{f_R})\frac{\gamma_5}{2}Z'{}^\mu\right) f~.
\label{Eq1}
\eeq
For the special case in which $q_{f_L}=-q_{f_R}$ the gauge boson is a pure axial vector. Many phenomenological studies contemplate a new axial vector which couples to Standard Model (SM) fermions. In particular, they are common in various scenarios for providing a portal between dark matter (DM) and SM states, e.g.~\cite{deSimone:2014pda,Hooper:2014fda,Lebedev:2014bba,Kahlhoefer:2015bea},  partially because if either the DM or SM fermions couple only axially to the vector mediator, the direct detection cross section is either spin-dependent or suppressed by factors of the DM velocity or momentum exchange.

Charging the SM fermions under a new U(1)$'$, in the absence of additional chiral fermions,  generically leads to the U(1)$'$ being anomalous. However, many studies neglect to specify the field content which would lead to an anomaly free theory \cite{deSimone:2014pda,Lebedev:2014bba,Kahlhoefer:2015bea,Buchmueller:2014yoa,Arcadi:2014lta,Chala:2015ama}.  Notably, any anomalous set of fermions can be embedded into a larger set which is anomaly free and whose members carry only rational charges  \cite{Batra:2005rh,Morrissey:2005uz}. Still, the associated extra states are typically charged under the SM gauge group.\footnote{
Cancellation mechanisms beyond new field content are avaliable in extra dimensional gauge theories, most prominently the Green-Schwarz mechanism \cite{Green:1984sg} and anomaly inflow \cite{Callan:1984sa}. For reviews see e.g.~\cite{Harvey:2005it,Scrucca:2004jn}.  Here we restrict our discussion to anomaly cancellation through new chiral fermions.}   As we will argue, these new states can not be arbitrarily separated from the mass scale of the axial vector. Thus it is important to consider the UV theory since such states are in principle observable at collider experiments.  Whilst there are many occurrences of complete anomaly free models of $Z'$ with general couplings in the literature  \cite{Langacker:2008yv}, there is a lack of examples for pure axial vectors.\footnote{Note that examples of axial vector models with anomaly cancelling exotics are presented in \cite{Hooper:2014fda}; model `Axial-A' is anomaly free, `Axial-B' is anomalous, and `Axial-Leptophobic' is anomaly free if one adds exotics $\psi^l_R$ and $\psi_L^e$ with U(1)$'$ charge zero.} Thus it is of interest to find anomaly free spectra for different scenarios in which the SM fermions interact with a new axial vector. 

We also note that new abelian gauge bosons are motivated from a GUT perspective, as large gauge groups naturally break to the Standard Model group supplemented with abelian factors. The breaking pattern may include U$(1)'$ factors and anomaly cancellation can be inherited from the matter content under the larger gauge group, as in the case of the {\bf 27} of $E_6$ under its axial subgroup U$(1)_\psi$~\cite{Langacker:2008yv}.
However, finding GUT completions for specific charge assignments can be challenging, and thus here we examine systematic `bottom-up' methods of anomaly cancellation without references to GUTs. Moreover, the GUT structure adds extra states not involved in anomaly cancellation and, to avoid proton decay, the U$(1)'$ scale is restricted to be near the GUT scale. Without requiring gauge coupling unification, by contrast, there is greater freedom in cancelling anomalies with new chiral exotics.

This paper is structured as follows: In Section~\ref{Sec2} we discuss the requirements for anomaly cancellation when the gauge structure of the SM is supplemented with a new U(1)$'$ factor, focusing on the case in which the U(1)$'$ gauge boson has only axial vector couplings to the SM fermions (and DM).
In Section~\ref{Sec3} we explore systematic methods for generating anomaly free models by adding new chiral fermions to the spectrum. Subsequently, we use these techniques to identify a number of anomaly free spectra for axial vector models of interest. Section~\ref{Sec4} considers the model building requirements for giving mass to the SM and exotic fermions.  In Section~\ref{Sec5} we ask at what scale the effective low energy description breaks down due to a loss of renormalisability, necessitating the introduction of new fermions, as well as the perturbativity bound on the U(1)$'$ coupling induced by the fermions. As one of the main motivations for these models is to use the axial vector as a portal to connect SM fermions and DM, Section~\ref{Sec6} considers the requirements for obtaining the observed DM relic density due to freeze-out via the axial vector, and the corresponding constraints from direct and indirect detection experiments and LHC searches.  Section~\ref{Sec7} presents some concluding remarks.  

Additional relevant content appears in the appendices. For completeness we give some anomaly free models for the case of a pure vector $Z'$ in \ref{ApC}.   We also show alternative sets of anomaly free sets of fermions with axial vector $Z'$ in \ref{ApB} and we give an explicit example of the algebraic constructions of anomaly free spectra in \ref{ApA}.


\section{Gauge Anomalies and Axial Vectors}
\label{Sec2}

Gauge symmetries are associated with conserved currents $\partial^\mu j_\mu=0$. In chiral gauge theories, unless the charges are appropriately arranged, anomalies from loop diagrams generically spoil gauge invariance $\partial^\mu j_\mu\neq0$. As is well known the SM anomaly conditions arise from triangle diagrams involving the following gauge interaction structures:
\begin{center}
SU(2)${}^2\times$U(1)${}_Y$,\hspace{1cm} SU(3)${}^2\times$U(1)${}_Y$,\hspace{1cm} ${\rm [Gravity]}^2\times$U(1)${}_Y$,\hspace{1cm}  U(1)${}_Y^3.$
\end{center}
The requirement that the anomaly coefficients satisfy $\mathcal{A}\propto\partial^\mu j_\mu=0$ for each of the triangle diagrams above gives rise to the following four conditions, respectively:
\beq
\mathcal{A}_{WWB}:=\sum_{f_L {\rm /w~SU(2)}} C_2[f_L] d_3[f_L] Y[f_L] &~- \sum_{f_R {\rm /w~SU(2)}}  C_2[f_R]d_3[f_R] Y[f_R] = 0~,\\[5pt]
\mathcal{A}_{ggB}:=\sum_{f_L {\rm /w~SU(3)}} C_2[f_L] d_2[f_L] Y[f_L] &~- \sum_{f_R {\rm /w~SU(3)}}  C_2[f_R]d_2[f_R] Y[f_R] = 0~,\\[5pt]
\mathcal{A}_{GGB}:=\sum_{f_L}d_2[f_L] d_3[f_L] Y[f_L] &~- \sum_{f_R}d_2[f_R] d_3[f_R] Y[f_R] = 0~,\\[5pt]
\mathcal{A}_{BBB}:=\sum_{f_L} d_2[f_L] d_3[f_L] (Y[f_L])^3 &~- \sum_{f_R} d_2[f_R] d_3[f_R] (Y[f_R])^3 = 0~,
\label{A1}
\eeq
where $d_N$ and $C_2$ are the dimension and quadratic Casimir of a given representation under SU($N$), and $Y$ is the hypercharge of a given state. The sums run over the left-handed (LH) and right-handed (RH) fermions respectively, and in the first/second condition the sum is restricted to representations of SU(2)/SU(3) only. Note that the other triangle diagrams cancel trivially.

There is also the Witten anomaly  \cite{Witten:1982fp} which places additional restrictions on field content transforming under groups which are equivalent to Sp($N$). In particular, an SU(2)${}_L\cong$ Sp(1) gauge theory with an odd number of LH-fermion doublets (and no other SU(2)-charged fermions) is inconsistent. However, as the SM is anomaly free and here we add fermions in vector-like pairs under the SM gauge group, or mimicking the SM generations, the Witten anomaly will not constrain our constructions. 
Furthermore, we restrict our field content to states with rational charges. This is motivated from charge quantisation considerations. In particular, this constraint plays a role in simple UV completions into larger GUT groups. Whilst no theorems forbid irrational charges in field theories, they are disfavoured in UV completions to GUTs \cite{Li:1981un}, and forbidden in quantum theories of gravity \cite{Banks:2010zn}.


\subsection{U(1)$'$ Anomaly Conditions}
\label{s2.1}

An extension of the SM gauge symmetry by an abelian factor, SU(3)$\times$SU(2)${}_L\times$U(1)${}_Y\times$U(1)$'$, introduces further anomaly conditions in addition to those of eq.~(\ref{A1}).  The vanishing of these new anomalies constrains the charges $z$ of states transforming under U(1)$'$, including any new fermions \cite{Barr:1986hj}. 
 First there are the four analogues to those involving U(1)${}_Y$, namely,
\begin{center}
 SU(2)${}^2\times$U(1)$'$,\hspace{1cm}  SU(3)${}^2\times$U(1)$'$,\hspace{1cm}  ${\rm [Gravity]}^2\times$U(1)$'$,\hspace{1cm}  U(1)$'{}^3$.
\end{center}
The coefficients $\mathcal{A}_{WWZ'}$, $\mathcal{A}_{ggZ'}$, $\mathcal{A}_{GGZ'}$, and $\mathcal{A}_{Z'Z'Z'}$ are direct analogues of eq.~(\ref{A1}) except with $Y$ replaced by $z$.
Two further anomaly conditions arise from mixed U(1)-U(1)$'$ diagrams 
\begin{center}
 U(1)${}_Y\times$U(1)$'{}^2$,
\hspace{1cm} U(1)$'\times$U(1)${}_Y^2$.
\end{center}
The associated anomalies vanish given the following conditions 
\beq
\mathcal{A}_{Z'Z'B}:=\sum_{f_L} d_2[f_L] d_3[f_L] Y[f_L] (z[f_L])^2 &~- \sum_{f_R} d_2[f_R] d_3[f_R] Y[f_R] (z[f_R])^2 = 0~,\\
\mathcal{A}_{BBZ'}:=\sum_{f_L} d_2[f_L] d_3[f_L] z[f_L] (Y[f_L])^2 &~- \sum_{f_R} d_2[f_R] d_3[f_R] z[f_R] (Y[f_R])^2 = 0~.
\eeq
There could also be an SU(3)${}^3$ anomaly with the addition of new chiral fermions. However, if the exotics are added in vector-like pairs under the SM group this vanishes automatically. In what follows, we will use the compact notation $z_X \equiv z[X]$ for a given field $X$.

\subsection{Coloured Exotics and Anomaly Free U(1)$'$ Extensions}
\label{2.2}

The case in which a U(1)$'$ gauge boson has only  axial couplings to the SM fermions is distinguished as it implies:
\beq
z_q^{(i)}:= z_Q^{(i)}=-z_u^{(i)}=-z_d^{(i)} \hspace{1cm} {\rm and} \hspace{1cm} z_l^{(i)}:= z_L^{(i)}=-z_e^{(i)}~.
\label{ax}
\eeq
The index $i=1,2,3$ denotes the SM generation.
Furthermore, if DM states $\chi_L$ and $\chi_R$ are present and couple axially to the $Z'$, it follows that
\beq
z_{\rm DM}:= z_{\chi_L}=-z_{\chi_R}.
\eeq 

Interestingly, the anomaly condition for SU(3)${}^{2}\times$U(1)$'$ alone immediately yields some useful information. Consider an axial vector which couples to quarks, thus $z_q^{(i)}\neq0$.  In the absence of new coloured states the SU(3)${}^{2}\times$U(1)$'$ anomaly $\mathcal{A}_{ggZ'}$ is
\beq 
\mathcal{A}_{ggZ'}= 2\left(z_q^{(1)}+z_q^{(2)}+z_q^{(3)}\right)~.
\label{ggz}
\eeq
In the case that the U(1)$'$ charge assignments for the SM fermions are mirrored in each generation ($z_q^{(1)}=z_q^{(2)}=z_q^{(3)}$), or only one generation is charged under U(1)$'$ (for instance $z_q^{(1)}=z_q^{(2)}=0$), then $\mathcal{A}_{ggZ'}$ will not vanish unless new coloured chiral fermions are introduced. Notably, the constraints from collider searches for coloured exotics are substantially more stringent than for uncoloured states. In the absence of new coloured fermions the anomaly condition of eq.~(\ref{ggz}) enforces
\beq
z_q^{(1)}+z_q^{(2)}+z_q^{(3)}=0~,
\eeq 
which requires different U(1)$'$ charges between generations of SM quarks. Allowing the U(1)$'$ charge assignments to differ between different generations introduces substantial freedom. In what follows we restrict ourselves to the cases where either the U(1)$'$ charges are replicated in the generation structure, or only one generation is charged under the U(1)$'$.

It is worth noting that in the pure vector case this anomaly cancels trivially, as when $z_Q^{(i)}=z_u^{(i)}=z_d^{(i)}$, $\mathcal{A}_{ggV}=0$ automatically without new coloured states. In \ref{ApC} we present some anomaly free models for the pure vector case, to illustrate that anomaly cancellation is typically much simpler in this scenario.


\section{Construction of Anomaly-Free Axial Vector Models}
\label{Sec3}

To calculate the anomaly coefficients one sums over all loops of chiral fermions, cf.~eq.~(\ref{A1}), including any chiral fermion exotics. Anomaly cancellation generically requires, and constrains, new exotic field content. For certain choices the exotic fermions automatically preserves the anomaly cancellation of the SM group. For instance, the exotics can mirror the SM fermion U(1)$'$ charges in order to cancel anomalies (Section~\ref{s3.1}). Alternatively, the exotics can constitute vector-like pairs under the SM gauge group, but have chiral charges under U(1)$'$ (Sections~\ref{s33} and \ref{s32}).  Moreover, with appropriate charges and representations one can cancel anomalies arising from diagrams involving the U(1)$'$ gauge bosons.

Whilst, in principle, one can introduce exotics in a variety of representations to arrange for anomaly cancellation, the most straightforward approach is to restrict the new field content to the fundamental representations of the SM group. Thus we restrict our analysis to the case that the exotics emulate the SM fermions, including hypercharge assignments (although this could be relaxed). We denote the new exotics as {\em primed} versions of their SM counterparts, and list them in Table \ref{Tab1}. In this section we will outline manners to systematically construct anomaly free sets of fermions. These techniques will be subsequently used in the construction of a selection of motivated scenarios of axial vector extensions of the SM.


\begin{table}[t!]
\begin{center}
\def\str{\vrule height11pt width0pt depth7pt}
\begin{tabular}{| c | c | c | c | c  | c |}
    \hline\str
    ~Field  Name~&~ U(1)${}_Y$ ~&~ SU(2)${}_L$~&~ SU(3)${}$~  \\[5pt]
     \hline\str
    ~~$Q_L^{i}$, $Q'_{L,R}$ & $1/3$ & 2 & 3
  \\[3pt]
    ~~$u_R^i$, $u'_{L,R}$  & $4/3$ & 1 & 3
    \\[3pt]
    ~~$d_R^i$, $d'_{L,R}$  & $-2/3$ & 1& 3
   \\[3pt]
    ~~$L_L^i$, $L'_{L,R}$  & $-1$ & 2 & $1$
 \\[3pt]
    ~~$e_R^i$, $e'_{L,R}$  & $-2$ & 1& 1 
    \\[3pt]
    ~~$\nu_R$,~${\chi}_{L,R}$ & 0 & 1 & 1 
   \\[3pt]
    ~~$H$ & 1 & 2 & 1 
     \\[3pt]
    \hline
\end{tabular}
\caption{The representation structure of the SM states, along with fermion exotics in matching representations. Here we assume the dark matter $\chi$ is a SM singlet; the $\nu_R$ entry indicates other singlets which do not constitute the dark matter. For the SM fields the index $i$ indicates the generation structure ($i=1,2,3$), there could also be multiple copies of any given exotic. The notation permits for an index $z$ for a U(1)$'$ charge and we will give anomaly free  assignments for $z$.} 
\label{Tab1}
\end{center}
\end{table}


\subsection{Mirror Constructions}
\label{s3.1}

In the case that the new exotics mirror the SM fields there is a simple manner to cancel any anomalies involving U(1)$'$ gauge bosons which we outline below. However, as we discuss in Section~\ref{Sec4}, this model requires a doubling of the exotics, or a non-minimal scalar sector in order to give masses to the anomaly cancelling fermions.

Each generation of the SM is an anomaly free set. However, if the SM fields are charged under the U(1)$'$, this introduces new anomaly contributions. Notably, anomaly  cancellation is automatic if for every SM fermion an exotic in the same representation of SU(2) and SU(3)  is introduced which either {\em i)}.~has the same U(1)${}_Y$ and U(1)$'$ charges but opposite chirality,   or {\em ii)}.~with matching chirality, but opposite U(1)${}_Y$ and U(1)$'$ charges. For instance, suppose that  $Q_L$ carries U(1)$'$ charge $z_q$, which we denote as $(3,2)_{\nicefrac{1}{3},z_q}$, one might add either a LH exotic in the representation $(3,2)_{-\nicefrac{1}{3},-z_q}$ or a RH exotic in  $(3,2)_{\nicefrac{1}{3},z_q}$. We call this approach the {\em mirror construction} for generating anomaly free sets of fermions.   
 
For each SM fermion, one adds a corresponding exotic. Therefore the mirror construction ensures that anomaly cancellation occurs state by state, and thus generation by generation. In the case that one adds opposite chirality mirror partners, then each exotic forms a vector-like pair with one of the SM fermions. If additional SM singlet states charged under U(1)$'$ are also introduced, such as DM fields $\chi_L$ and $\chi_R$, the contributions from these states can be cancelled via the addition of RH neutrino states $\nu_R$ with appropriate U(1)$'$ charges. Note that SM fermions that do not carry U(1)$'$ charges must still have exotic partners (with U(1)$'$ charge zero) to cancel the anomalies of the SM gauge group, unless the set of states uncharged under U(1)$'$ have the correct representations to fill out a full SM generation.

If only certain SM fermions carry U(1)$'$ charges, such as a single generation, then such mirror constructions have relatively minimal fermion spectra. However, if all or many SM fermions carry U(1)$'$ charges then, it implies the introduction of a large number of exotics. Note that there are generically flavour constraints on non-universal $Z'$ models, which are somewhat alleviated in the case that the first two generations have the same U$(1)'$ charge~\cite{Langacker:2000ju}.

In the rightmost two columns of Table \ref{Tab5} we show two examples in which a single SM generation is charged under U(1)$'$ and the anomalies are cancelled through mirror exotics.  In the remainder of this section we consider more general algebraic approaches which can present smaller anomaly free sets of fermions.

\begin{table}[H]
\begin{center}
\def\str{\vrule height11pt width0pt depth5pt}
\begin{tabular}{| c | c | c | c | c  | c | c|}
    \hline\str
    ~~~Field ~~~&~~~ $\sharp1$ ~~~&~~~ $\sharp2$~~~&~~~ $\sharp3$~~~ &~~~$\sharp4$ ~~~ &~~~ $\sharp5$~~~&~~~ $\sharp6$~~~ \\[2pt]
     \hline\str
    $z[Q_L]$          & 1       &  1      & 1      &  0   & 1      &  1  \\
    $z[u_R]$          &  -1      &  -1     & -1      &  0 & -1      &  -1 \\
    $z[d_R]$          & -1     &  -1       & -1     &  0  & -1     &  -1       \\
    $z[L_L]$          &  1     &    1      & 0      &  1   &    1    & 0 \\
    $z[e_R]$          &  -1     &  -1     &0      & -1    &  -1     & 0  \\
    $z[{\chi}_{L}]$   &    -      &  9    &  9    &  -9/4   &   1    &  1   \\
    $z[{\chi}_{R}]$   &    -      &  -9  &  -9    & 9/4   &   -1   &  -1  \\
            \hline\str
    $z[Q'_{L}]$        &   1      &   1     &   1   &  -     &   -   &  -    \\
    $z[Q'_{R}]$       &   3      &   -1    &   0    &  -   &    1     &  1  \\
    $z[u'_{L}]$         &   -3    &   -2    &  -2     &  -2   &   -1    &  -1 \\
    $z[u'_{R}]$        &   4     &    3   &  -1     &  5/2    &    -    & -      \\
    $z[d'_{L}]$         &  3      & -6     & -2     &  2   &     -1   &  -1  \\
    $z[d'_{R}]$        &   4     & 5     & 11       &  -5/2  &   -    & -   \\
    $z[L'_{L}]$         &  -9     & -82/3    &    -49/12    & -157/48   &   -    & -     \\
    $z[L'_{R}]$         & -3     & -28/3   &  95/12    &  -13/48   &    1    &   0     \\
    $z[e'_{L}]$          &  -13  &  -100/3    &   103/6  &  -85/24   &  -1        &  0  \\
    $z[e'_{R}]$         &    -16 & -127/3  &  67/6    &  -121/24   &   -     & -   \\
        \hline\str
    $z[\nu_R]$   &  -   &  -   & -   &  -  &  1  &   1       \\
    $N[\nu_R]$   &  -   &  -   & -   &  -  &  2   &   2        \\
        \hline\str
    $b_{m_z}$   &   45    &  207  & 198   &  153/8 &  17  &  14       \\[1.5pt]
    $b_{m_z}+b_M$   &  860    &  15038/3   & 14065/12   & 90697/192 &  34  &   28      \\[1.5pt]
    $\mathcal{A}^{\rm SM}_{Z'Z'Z'}+\mathcal{A}^{\rm DM}_{Z'Z'Z'}$   &  45   &  45+1458   & 36+1458    & $9-729/32$  &  15+2 &   12+2       \\[1.5pt]
     \hline
\end{tabular}
\caption{Charge assignments $z[f]$ and multiplicities $N[f]$ of states which give anomaly free spectra. For Models $\sharp$1-$\sharp$4 the U(1)$'$ charges are mirrored in each SM generation. In Models $\sharp$5-$\sharp$6, only one generation carries  U(1)$'$ charge. A dash `-' indicates that the corresponding state is absent in a given model. See Table \ref{Tab1} for representations and charge assignments of states under the SM gauge group. We also give $b_{m_Z}$ and $(A^{\rm SM}_{Z'Z'Z'}+\mathcal{A}^{\rm DM}_{Z'Z'Z'})$, the $\beta$-function and U(1)${}^3$ anomaly contributions from the SM fermions plus DM, and $b_M$ the exotics $\beta$-function contribution, which are referenced in Section~\ref{Sec5}.
}
\label{Tab5}
\end{center}
\end{table}
\afterpage{\clearpage}
\newpage

\subsection{An Algebraic Construction}
\label{s33}

Requiring anomaly cancellation gives a set of equations, which for a definite set of charges can be solved directly. Specifically, consider the case that the SM gauge group is extended by a U(1)$'$ factor, with no additional states except for those required to cancel anomalies and that all of the SM fermions couple to the gauge boson with only axial vector couplings, thus the charges satisfy eq.~(\ref{ax}). Further, we assume the charges are the same for each generation: $z_q^{(1)}=z_q^{(2)}=z_q^{(3)}$ (similarly for leptons). To simplify the Higgs sector required to give Yukawa couplings to the SM fermions (as we discuss in Section~\ref{s4.1}) we also take  $z_q^{(i)}=z_l^{(i)}$, for all generations $i$.\footnote{We refer to this scenario as `Model $\sharp$1' in later sections.} To emphasise the relation between charges we write   $z_{\rm SM}:=z_q=z_l$.

We will assume that the anomaly cancelling exotic fermions form a single full generation of vector-like fermions under the SM group $Q'_L,~Q'_R,~u'_L,~u'_R,~d'_L,~d'_R,$ $l'_L,~l'_R,$ $e'_L,~e'_R,$ which mimic their SM namesakes (see Table \ref{Tab1} for definitions of the representations). That this set of fermions is vector-like under the SM group implies that the SM chiral anomalies and the Witten anomaly are resolved automatically.
 Interestingly in this case the equations which ensure anomaly cancellation can be solved directly to arrive at a general, unique set of seven conditions which generically determine anomaly free sets of fermions with rational charges:\footnote{Other possible solution sets generically yield irrational charges, which are theoretically disfavoured \cite{Li:1981un,Banks:2010zn}.}
\beq
\vspace{5mm}
& 
 z_{Q'_R} = z_{Q'_L} + 2 z_{\rm SM}, \hspace{15mm}
z_{u'_R} = 7 z_{\rm SM} + z_{u'_L}, \\[1pt]
&
z_{d'_R} = z_{d'_L} + z_{\rm SM},  \hspace{15mm}
   z_{L'_R} = z_{L'_L} + 6 z_{\rm SM}, 
\\[1pt]
  z_{e'_L} = \frac{1}{3} (z_{d'_L} + & 6 z_{L'_L} - 2 z_{Q'_L} - 28 z_{\rm SM} - 14 z_{u'_L}), 
 \hspace{10mm} 
z_{e_{R'}}=z_{e_{L'}} -3 z_{\rm SM}\\
  z_{L'_L} = \frac{1}{\Omega}\Big(-8 z_{d'_L}^2& - 4 z_{d'_L} z_{Q'_L} - 32 z_{Q'_L}^2 - 74 z_{d'_L} z_{\rm SM} + 
      58 z_{Q'_L} z_{\rm SM} \\ & - 404 z_{\rm SM}^2 - 28 z_{d'_L} z_{u'_L} + 56 z_{Q'_L} z_{u'_L} + 
      469 z_{\rm SM} z_{u'_L} + 133 z_{u'_L}^2\Big)~,
\eeq
with $\Omega=606 z_{\rm SM}+168 z_{u'_L} -12 z_{d'_L} + 24 z_{Q'_L}\neq0$.

The above set of equations uniquely characterises the solution set. Since the charges are all related through anomaly cancellation, fixing a subset of the charges determines the remaining charges; e.g.~taking $z_{\rm SM}=z_{Q'_L}=1$ and $z_{u'_L}= -z_{d'_L} = -3$ one obtains that for anomaly cancellation the other charges are required to be 
\beq
&\hspace{14mm} z_{Q'_R} = 3,\hspace{14mm} 
z_{u'_R} = 4 \hspace{15mm}
z_{d'_R} = 4, \\
z_{L'_L} &= -9, \hspace{12mm}
z_{L'_R} = -3, \hspace{12mm}
z_{e'_L} = -13, \hspace{12mm}
z_{e'_R}= -16.
\eeq
If any LH-RH pair obtains the same U(1)$'$ charge, the states are redundant for anomaly cancellation, and thus can be removed from the model if desired.

For models which also include fermion DM $\chi_L$ and  $\chi_R$ with U(1)$'$ charges which also have only axial couplings to the U(1)$'$ gauge boson, the situation is somewhat different. The additional freedom, due to the undetermined charge assignment of $z_{\rm DM}$, means that solving the conditions for anomaly cancellation with a single full generation of SM-vector-like fermions leads to six sets of solutions which each provide anomaly free spectra with rational charges. This is in contrast to the unique set found without the inclusion of DM. 
In this case, if one removes the $u'_L$ and $u'_R$ exotics (or make them vector-like under U(1)$'$ such that $z_{u'_L}=z_{u'_R}$) then again there is a unique set of equations which determine the anomaly free sets of fermions. However, it is useful to use a full generation of SM-vector-like fermions as this makes it easier to find anomaly free models with simpler charge assignments, avoiding fractional charges with large numerators and denominators. For brevity, we neglect to give the sets of equations which ensure anomaly cancellation with the addition of DM, but these can readily derived using {\em Mathematica} \cite{mathematica} or an analogous equation solver.

\subsection{General Algebraic Constructions}
\label{s32}

A more general approach to finding anomaly free sets of fermions with arbitrary charges was outlined in the work of Batra, Dobrescu and Spivak \cite{Batra:2005rh}, providing algebraic expressions for the U(1)$'$ charge assignments of the exotics and multiplicities of the SM singlets, as a function of the U(1)$'$ charges of the SM fermions.  Indeed, using this method one can systematically embed any anomalous set of fermions into a larger theory which is anomaly free and where the fermions carry only rational charges.

To systematically find anomaly free spectra for SM fermions with arbitrary charges under U(1)$'$, one should introduce at least one chiral pair of states transforming under SU(3), one chiral pair transforming under SU(2) and one chiral pair charged under hypercharge. This set of exotics provides sufficient freedom to cancel the anomalies arising from the diagrams between mixed U(1)$'$ and SM gauge bosons. Following   \cite{Batra:2005rh}, we introduce pairs of exotics $d'_L,~d'_R,$ $L'_L,~L'_R,$ $e'_L,~e'_R$, which are vector-like under the SM group. (Note that, unlike the previous sections, we do not introduce $Q'$ or $u'$ exotics here.)

Firstly, from the requirement of vanishing anomalies for the three diagrams involving two SM gauge bosons (U(1)${}_Y^2\times$U(1)$'$, SU(2)${}^2\times$U(1)$'$, SU(3)${}^2\times$U(1)$'$),
one can readily obtain equations for the difference between the charges of the LH and RH exotics, i.e.~$(z_{d'_L}-z_{d'_R})$, $(z_{L'_L}-z_{L'_R})$, $(z_{e'_L}-z_{e'_R})$. The next step in the construction is to posit a basis for the sum of the exotic charges in terms of a linear combination of the U(1)$'$ charges of the SM fields:
\beq
(z_{X'_L}+z_{X'_R})=C^X_{1}z_q+C^X_{2}z_{\rm DM}+C^X_{3}z_l
\hspace{15mm} {\rm for}~X=d,L,e~.
\label{lin}
\eeq
Given the difference of the charges of the LH and RH exotics, and the above form of the sum of these charges, one can  take linear combinations of these equations to obtain expressions for the U(1)$'$ charges of each of the exotics in terms of the SM charges and the constants $C^X_{i}$. Then demanding the vanishing of the U(1)$'{}^2\times$U(1)${}_Y$ anomaly for arbitrary SM U(1)$'$ charge assignments leads to relations between the various constants $C^X_{i}$. This typically leaves a number of constants undetermined.

It remains to arrange for the ${\rm [Gravity]}^2\times$U(1)$'$, and U(1)$'{}^3$ anomalies to vanish. We assume the spectrum contains two types of RH neutrinos $N_1\times\nu_R^{(1)}$ and $N_2\times\nu_R^{(2)}$ which are SM singlets with U(1)$'$ charges $z[\nu_R^{(1)}]=-1$ and $z[\nu_R^{(2)}]=2$, and $N_\alpha$ indicate the number of copies of these states. Then insisting that the two remaining anomalies vanish, one obtains an equation for multiplicities $N_\alpha$ of the RH neutrinos states. If $N_\alpha<0$ this implies that $|N_\alpha|$ RH (or LH) SM singlets with charge $z[-\nu^{(\alpha)}_R]$ (with charge $z[\nu^{(\alpha)}_R]$) are required.

Then fixing the SM charges and the undetermined $C^X_{i}$, any choice which yields integer values for $N_1$ and $N_2$ gives a consistent anomaly free fermion spectrum.  This commonly leads to high multiplicities $N_1$ and $N_2$. However, following the procedures outlined in   \cite{Batra:2005rh},  the number of SM singlets can often be replaced with a smaller set of RH neutrinos with larger U(1)$'$ charges.  \ref{ApB} gives anomaly free sets of fermions for various models which are derived via an application of the method of  \cite{Batra:2005rh}. Additionally, in \ref{ApA} we present an explicit derivation using this method.


\subsection{A Selection of Axial Vector Models}
\label{s3.4}

There are many scenarios involving axial vectors which could be of interest. Here, we highlight a number of motivated extensions of the SM here and construct anomaly free spectra which realise these scenarios. Specifically, we will consider the following cases: 

\begin{enumerate}

\item[{\bf Model $\sharp1$.}]  The simplest scenario is the extension of the SM gauge group with an additional U(1)$'$ factor, where all of the SM fermions couple axially to the $Z'$, and the U(1)$'$ charge assignments of the SM fermions are replicated in the generation structure.\footnote{Model $\sharp1$ is also relevant when including scalar DM or fermion DM with vector couplings to $Z'$.}
 
\item[{\bf Model $\sharp2$.}]  A minimal extension of Model $\sharp1$ is to include chiral fermion DM states which are SM singlets, and also couple axially to the gauge boson of U(1)$'$. We shall also assume the scenario of fermion DM charged under U(1)$'$ in Models $\sharp3$-$\sharp6$.

  \item[{\bf Model $\sharp3$.}]   A slight modification to  Model $\sharp2$ is the case the axial vector has no tree level couplings to leptons by enforcing $z_l=0$, thus yielding a leptophobic axial vector.
  
\item[{\bf Model $\sharp4$.}]   Conversely, one might consider a leptophilic case with $z_l\neq0$ and $z_q=0$.

\item[{\bf Model $\sharp5$.}]  Not all  SM generations need be charged under U(1)$'$ and we consider the case that only a single generation (1G)  has U(1)$'$ charges. For example $z^{(1)}_{q}=z^{(1)}_{l}=z^{(2)}_{q}=z^{(2)}_{l}=0$.

\item[{\bf Model $\sharp6$.}]  Moreover, it could be that only a small subset of SM fermions carry U(1)$'$ charge. Specifically, we consider the case that only $z_Q^{(3)}=-z_{u}^{(3)}=-z_d^{(3)}\neq0$,  with all other SM fields neutral under U(1)$'$. This realises a single generation leptophobic model.

\end{enumerate}

A summary of the above models is given in Table \ref{Tab2}. In Table \ref{Tab5} we present anomaly free sets of fermions which realise Models $\sharp1$-$\sharp6$ outlined above. The anomaly free sets presented for Models $\sharp$1-$\sharp$4 are generated via the method of Section~\ref{s33}, whilst the spectra for Models  $\sharp$5 \& $\sharp$6 come from the mirror construction, as discussed in Section~\ref{s3.1}. Alternative anomaly free sets for Models $\sharp$1-$\sharp$6  which eliminate some of the coloured exotics at the price of introducing RH neutrinos, as discussed in Section~\ref{s32}, are given in \ref{ApB}. 



\begin{table}[t!]
\begin{center}
\def\str{\vrule height11pt width0pt depth7pt}
\begin{tabular}{| l | c | c | c | c  | c | c |}
    \hline\str
   Name ~&~$n_G$ ~ & Lepto-phobic/philic? \\[2pt]
     \hline\str
    $\sharp1$.~Universal Model &~~3    & \xmark\\[2pt]
   $\sharp2$.~/w DM Model &~~3     & \xmark\\[2pt]
    $\sharp3$.~$L$-phobic~Model  &~~3     & Leptophobic\\[2pt]
    $\sharp4$.~$L$-philic~Model  &~~3    & Leptophilic\\[2pt]
    $\sharp5$.~1G-Model  &~~1 & N/A     \\[2pt]
    $\sharp6$.~$t$-$b$-Model  &~~1     & Leptophobic\\[2pt]
       \hline
\end{tabular}
\caption{Summary of the models we study. $n_G$ is the number of SM generations charged under U(1)$'$.}
\label{Tab2}
\end{center}
\end{table}

\section{Mass Generation}
\label{Sec4}

We next consider what form of scalar sector is required to give masses to the SM and exotic fermions for the axial vector models outlined in the previous section. These considerations are often absent in phenomenological studies, but regularly require non-trivial model building. We do not attempt to be comprehensive, but rather make some general remarks.

\subsection{Mass Generation for Standard Model Fermions}
\label{s4.1}

If all SM fermions couple axially to U(1)$'$, then gauge invariance forbids a full set of  SM Yukawa couplings from a single Higgs. The reason is that the U(1)$'$ charge of the bilinears is $z[\bar Q_L u_R]=z[\bar Q_L d_R]=2z_q$ for axial vector couplings. To form a gauge invariant operator  $H^\dagger \bar Q_L u_R$ requires $z[H^\dagger]=-2z_q$, but this forbids the Yukawa couplings for the down-quarks and leptons since in the SM these involve the conjugate field. This difference in SM fermion bilinears is even more apparent if only some generations are charged under U(1)$'$. Finally, electroweak precision data also constrains the U(1)$'$ charge of the SM Higgs because of the induced $Z$-$Z'$ mixing.
The remaining mass terms could still arise  via renormalisable terms involving additional Higgses, as in a Type II Two Higgs Doublet Model \cite{Branco:2011iw}, or due to higher dimension operators. Perhaps the simplest manner to give masses to all of the SM fermions is for the Higgs to be uncharged under U(1)$'$ and introduce a scalar $S$ which is charged under U(1)$'$, but is a SM singlet, such that there are dimension five effective operators for the remaining SM fermions: $(1/\Lambda_*)SH^{\dagger}\bar Q_Lu_R$, etc. This operator is generated by physics integrated out at the scale $\Lambda_*$, and the theory must UV complete to a renormalisable Lagrangian at energies approaching $\Lambda_*$. This is reminiscent of the Froggatt-Nielsen mechanism  \cite{Froggatt:1978nt}.

Since $S$ is a SM singlet, gauge invariant dimension five operators can be formed using $S$ and $S^{\dagger}$, which give mass terms to all SM fermions once $S$ acquires a VEV $\langle S \rangle$. The $\langle S \rangle$ breaks the U(1)$'$, and thus the fermion masses are connected to the axial vector mass. This scenario is no longer UV complete, and one expects additional states to enter at the scale which generates the higher dimension operators, which could be near the TeV scale.  For mass terms induced due to $\langle S \rangle^n$ this yields effective Yukawa couplings of order $(\langle S\rangle/\Lambda_*)^n$. However, a good effective field theory (EFT) requires $\langle S\rangle\lesssim \Lambda_*$, and thus it is challenging to obtain $\mathcal{O}(1)$ Yukawa couplings via high dimension operators. Hence, from a model building stance, the use of high dimension operators to generate the top Yukawa is disfavoured. 

In an EFT with a $Z'$, where the scalar $S$ responsible for breaking U(1)$'$ has been integrated out, the VEV of this scalar $\langle S\rangle\equiv v'$ introduces an order parameter, which acts as a cutoff of the EFT. The VEV responsible for breaking U(1)$'$ generates the $Z'$ mass $m_{Z'}\simeq g'v'$, and the mass of the associated scalar is parametrically $m_S\sim \lambda_S v'$, where $\lambda_S$ is the $S$ quartic coupling.
Unitarity of the EFT describing the light SM fermions $f$, DM, and $Z'$ requires that $m_f,~m_{\rm DM}\lesssim \frac{m_{Z'}}{g'}\simeq v'$  and the bosons should satisfy $m_{Z'}, m_S\lesssim v'$; see e.g.~\cite{Kahlhoefer:2015bea} for further discussion. For example, giving the top a U(1)$'$ coupling $g'\sim1$ would imply a lower bound on the $Z'$ mass of $m_{Z'}\gtrsim 175$ GeV. This bound is stronger for heavy DM states
\beq
m_{Z'}\gtrsim1~{\rm TeV}\left(\frac{g'}{1}\right)\left(\frac{m_{\rm DM}}{1~{\rm TeV}}\right)~.
\label{EFT}
\eeq

\begin{table}[b!]
\begin{center}
\def\str{\vrule height11pt width0pt depth5pt}
\begin{tabular}{| c | c | c | c | c  | }
    \hline\str
    ~~~Bilinear ~~~&~~~ $\sharp1$ ~~~&~~~ $\sharp2$~~~&~~~ $\sharp3$~~~ &~~~$\sharp4$ ~~~ \\[2pt]
     \hline\str
   SM $\overline{\rm LH}$-RH bilinears (e.g.~$z[\bar Q_Lu_R]$)                 &   -2  & -2 & -2 & -2                    \\
        \hline\str
  $z[\bar {\chi}_{L}{\chi}_{R}]$ &  -  &   - 18    &  -18    &  9/2       \\[1.5pt]
    $z[\bar Q'_{L} Q'_{R} ]$    &   2  &   -2     &  -1   &  -         \\[1.5pt]
    $z[\bar u'_{L} u'_{R} ]$     &   7   &   5    &  1     &  9/2       \\[1.5pt]
    $z[\bar d'_{L} d'_{R}]$      &  1    &    11   & 13     &  -9/2        \\[1.5pt]
    $z[\bar L'_{L}L'_{R}]$       &  6   &   18    &    12   & 3    \\[1.5pt]
    $z[\bar e'_{L}e'_{R}]$       & -3    &   -9     &   -6  &  -3/2        \\[1.5pt]
            \hline\str
   $z[\bar Q'_{L} u'_{R} ]$   & 3    &    2    &   -2   &   -       \\[1.5pt]
   $z[\bar Q'_{L} d'_{R} ]$   &  3  &   4     &   10   &   -       \\[1.5pt]
   $z[\bar Q'_{R} u'_{L} ]$   &  -6   &  -1     &   2  &    -      \\[1.5pt]
   $z[\bar Q'_{R} d'_{L} ]$   & 0   &   -5    &   -2  &    -      \\[1.5pt]
   $z[\bar L'_{L}  e'_{R} ]$   & -7   &   -15    &  61/4    & -85/48         \\[1.5pt]
   $z[\bar L'_{R}  e'_{L} ]$   & -10  &   -24     &   37/4   &    -157/48    \\[1.5pt]

            \hline\str
    No.~scalars for Yukawa terms & 5 & 5 &    4 &  3           \\[1.5pt]
      Min.~number of scalars   &  2     &  3    &  2    &  2     \\[1.5pt]
     \hline
\end{tabular}
\vspace{5mm}
\caption{Charges of fermion bilinears for Models $\sharp1$-$\sharp4$. Also shown is the number of exotic scalars needed to give vector-like masses to all exotics after VEV insertions (not including SM Higgs), and to give masses to all exotics via a combination of renormalisable and non-renormalisable operators with mass dimension six or less. These models need multiple scalars to give all fermions masses.  \label{Tab7}}
\end{center}
\end{table}

\newpage
\subsection{Mass Generation for Pairs of Exotic Fermions}
\label{s4.2}

Anomaly cancellation in models with axial vector $Z'$ requires an array of exotics with chiral charge under U(1)$'$, and, as can be seen from Table \ref{Tab5}, the pattern of U(1)$'$  charge assignments of these exotics is often complicated. As such these new fermions can not typically be given dimension four Yukawa couplings involving the Standard Model Higgs. The simplest manner to give masses to the exotic fermions is through the introduction of {\em exotic Higgses}, new SM singlet scalars charged under U(1)$'$ which acquire VEVs, and give masses to the various exotics via renomalisable interactions. In the case where the exotics come in pairs that are vector-like under the SM, they may acquire masses through renormalisable interactions involving the exotic Higgses. Due to the different U(1)$'$ charges of the new fermions, this generally requires one exotic Higgs for each exotic fermion pair.
In addition to the LH-RH mass bilinears (e.g.~$\bar Q_L' Q_R'$), the exotics could also have chiral mass bilinears (e.g.~$\bar Q_L' u_R'$). However, exotic fermion mass operators using these bilinears must include the SM Higgs field as well as an exotic Higgs, and are thus non-renormalisable.
An alternative approach is to add  fewer exotic Higgses, such that some of the exotic fermions do not have renormalisable mass terms, but higher dimension operators respecting the gauge symmetries can give masses after VEV insertions.

When the exotics acquire mass through U(1)$'$-breaking VEVs at the scale $\sim v'$, we expect that the masses of the gauge boson ($m_{Z'}\sim gv'$) and the exotics  ($M\sim y' v'$) should be comparable: $m_{Z'}\sim M$. Any hierarchical splitting of $m_{Z'}$ or $M$ from $v'$ arises primarily due to couplings. Moreover, perturbativity of $y'$ implies
\beq
M\lesssim v'\simeq \frac{m_{Z'}}{g'(m_{Z'})}.
\label{limv}
\eeq Thus separating the $Z'$ from the exotics requires a tuning of the Yukawa couplings such that $g'\ll y'$. In addition, the exotic fermions can not be made significantly heavier than $v'$.

\vspace{2mm}
\subsection{The Scalar Sector of Model $\sharp1$}
\label{s4.3}
\vspace{2mm}

Let us consider a specific example. Below, we outline a scalar sector for Model $\sharp1$ which can give masses to all of the SM fermions and chiral exotics.  To understand the charges required for exotic Higgses we should look at the net charge of the bilinear operators involving chiral exotic pairs. We give these for Models $\sharp$1-$\sharp$4 in Table \ref{Tab7}. In Model $\sharp$1, observe that most of these bilinears have different net charge, and thus five different scalars (with $|z|=1,2,3,6,7$) are required for these states to acquire vector-like mass terms via renormalisable Yukawa terms with VEV insertions.  We denote by $S_q$ an exotic Higgs with $z[S_q]=q$.

Note that one could replace terms involving $S_{-6}$ with non-renormalisable terms involving $S_{-2}$ and $S_{-3}$. For example, instead of $S_{-6} \bar L'_{L} L'_{R}$ mass terms can also arise from
\beq
\left[\frac{c_1}{\Lambda_*}\langle S_{-3}\rangle^2+
\frac{c_2}{\Lambda_*^2}\langle S_{-2}\rangle^3 
\right]\bar L'_{L} L'_{R}~,
\eeq
where $\Lambda_*$ is the cutoff of the EFT.
The operator involving $S_{-3}^2$ is dimension five, while the others are dimension six. However, if its coefficient is small, $c_1\ll c_2$, or if $\langle S_{-3}\rangle\ll\langle S_{-2}\rangle$, this operator will not necessarily dominate. Thus the number of exotic Higgses required can be reduced (cf.~Table \ref{Tab7}), but at the expense of UV completeness.

Now let us consider an example Lagrangian for the scalar sector of Model $\sharp$1. Suppose the SM Higgs $H$ has charge  $z[H]=-2$ and introduce two SM singlet scalars $S_1$ and $S_4$ with U(1)$'$ charges $z[S_1]=1$ and $z[S_4]=4$. With these states the SM Yukawa couplings can be constructed with a renormalisable interaction for the up-like quarks (useful for obtaining the large top Yukawa) and dimension five operators responsible for the down and lepton Yukawas 
\beq
\mathcal{L_{\rm SM}}
&\supset y_u^iH^\dagger \bar Q_Lu_R+\frac{y_d^i}{\Lambda_*}S_4H \bar Q_Ld_R+\frac{y_l^i}{\Lambda_*}S_4H \bar  L_Le_R
+{\rm h.c.}
\label{SMY}
\eeq
The scale suppression of the higher dimension operators can help realise the fermion hierarchy, as in the Froggatt-Nielsen mechanism \cite{Froggatt:1978nt}.
For the exotic fermions one can obtain vector-like masses via gauge invariant Yukawa terms involving the SM singlet scalars,
\beq
\mathcal{L_{\rm Ex}}
&\supset \frac{y_{Q'}}{\Lambda_*} {S^\dagger_1}^2 \bar Q'_{L} Q'_{R} 
+\frac{y_{u'}}{\Lambda_*^2}{S^\dagger_4}^2 S_1 \bar u'_{L} u'_{R} 
+y_{d'}S^\dagger_1 \bar d'_{L}  d'_{R} + \frac{y_{L'}}{\Lambda_*^2}S^\dagger_4 {S^\dagger_1}^2  \bar  L'_{L}L'_{R}
+ \frac{y_{e'}}{\Lambda_*}S_4S_1^\dagger \bar e'_{L} e'_{R}
+\cdots
\eeq
None of the leading SM mass terms involve $S_1$, and in contradistinction all of the exotic fermion mass terms involve $S_1$. Thus the magnitude of the $\langle S_1 \rangle$  is not restricted by the requirement that one reproduces the SM fermion masses and a large $S_1$ VEV can be used to decouple the exotic fermions. This results in a hierarchy between the exotic fermions, but we will not discuss this here. Moreover, a large $\langle S_1 \rangle$ breaks U(1)$'$ at a high scale, allowing for a $Z'$ which is much heavier than the weak scale. This avoids electroweak precision constraints from tree-level $Z$-$Z'$ mixing \cite{Erler:2009jh,Babu:1997st,Kahlhoefer:2015bea,Agashe:2014kda}, which for $m_{Z'} \gg m_Z$ require $m_{Z'} \gtrsim g' (14~\mathrm{TeV})$.

There are also mass terms from chiral bilinears, such as  $\bar Q_L'u_R'$, which must be paired with a Higgs $H$ field for SU(2) invariance and a combination of $S_1$ and $S_4$ fields to conserve U(1)$'$ charge. As discussed above, mass operators containing these bilinears, e.g. $S_4^\dagger S_1 H^\dagger\bar Q_L' u_R'$, are non-renormalisable, but can affect the mass splittings between exotic fermions.

Giving mass to certain fields via higher dimension operators implies that the EFT should break down around $\Lambda_*$, and one might ask what manner of physics can give rise to such operators. As an example, consider the dimension five operator $S_4H \bar  L_Le_R$ in eq.~(\ref{SMY}) which is responsible for the electron mass. This operator can arise from a vector-like pair of fermions $\psi_L$, $\psi_R$ in the representation $(1,1)_{-2,3}$ entering in the Lagrangian of the UV theory
\beq
L_{\rm UV}\supset y_\psi H \bar L_L \psi_L + y_\psi' S_4 \bar \psi_L e_R + m_\psi \bar \psi_L \psi_R +{\rm h.c.}
\label{uv}
\eeq
After integrating out $\psi$, one recovers the contact operator which gives mass to the electrons and the EFT cutoff can identified as $\Lambda_*=\frac{m_\psi}{y_\psi y_\psi'}$. While the introduction of high dimension operators necessitates new physics (for instance new fermions) in the UV theory, these states could be significantly above the weak scale.

\subsection{Mass Generation for Exotic Mirror Fermions}

In anomaly free models arising from mirror constructions, such as Models $\sharp$5 \& $\sharp$6, the situation is somewhat different. Since the exotic fermions are not introduced in pairs that are vector-like under the SM, but rather as copies of SM generations, one requires a new scalar which is a doublet under SU(2)${}_L$  to construct renormalisable Yukawa terms. Since the VEV of such a scalar breaks electroweak symmetry, it is constrained by electroweak precision and Higgs measurements. Moreover, in this case the exotics can not be much above the weak scale. Viable exotics require Yukawas near the perturbative limit, which implies new physics at the TeV scale, and thus such scenarios will be generically constrained by collider searches.

Alternatively, we may introduce further exotic fermions which do not disrupt the anomaly cancellation, and then give masses to the exotics in the same fashion as in Section~\ref{s4.2}. This can be achieved if one supplements the mirror constructions, such as Models $\sharp$5 \& $\sharp6$, with a full set of states with identical SM representations, zero U(1)$'$ charge, and opposite chirality to the existing exotic fermions. For example, for Model $\sharp5$ one would add $Q_L'$, $u_R'$, $d_R'$, $L_L'$, $e_R'$ with $z[Q'_L]=z[u'_R]=z[d'_R]=z[L_L']=z[e_R']=0$. Since they are uncharged under U(1)$'$, they obviously do not contribute to any anomalies involving U(1)$'$. Furthermore, since these states mimic an entire generation of SM fermions, and the anomalies in the SM cancel generation by generation, it follows that this spectrum is anomaly free.

The benefit of doubling the number of exotics is that now one can form Yukawa terms for the anomaly cancelling exotics which give vector-like masses after VEV insertions, similar to Section~\ref{s4.2}. For a given anomaly cancelling RH exotic $X_R$ one can form a LH-RH bilinear which has net charge $z[\bar X_L  X_R]=z[X_R]$ (similarly for LH anomaly cancelling exotics). For Models $\sharp$5 \& $\sharp6$, supplemented by a generation with opposite chirality and zero U(1)$'$ charge, all of the exotic $\overline{\rm LH}$-RH bilinears have net charge $1$. Thus one can give mass to all of the exotic fermions through a single new SM singlet scalar field $S$ with $z[S]=-1$ 
\beq
\mathcal{L}_{\rm Mir}\supset  S \bar Q_L'  Q_R' +S \bar u'_L  u'_R + S \bar d'_L  d'_R + S \bar L'_L  L'_R + S \bar e'_L  e'_R~.
\eeq
This model has a minimal scalar sector, and is UV complete, at the price of doubling the fermion content of the theory.


\section{Breakdown of Low Energy Theories}
\label{Sec5}

In this section we examine at what scale new physics is needed to mitigate a breakdown in the low energy theory, either due to a loss of renormalisability from uncancelled anomalies or, after introducing new fermions for anomaly cancellation, due to a loss of perturbativity of the U(1)$'$ gauge coupling $g'$.

\subsection{The Non-Perturbative Limit}

In the SM the hypercharges of fields are all $\mathcal{O}(1)$ and as a result the gauge coupling remains perturbative well beyond the Planck scale. However, as can be seen from Table \ref{Tab5}, the exotics required for anomaly cancellation in axial vector extensions of the SM often carry large  U(1)$'$  charges. As a result the U(1)$'$ gauge coupling $g'$ may quickly run non-perturbative. Indeed, shortly after the coupling nears the non-perturbative limit one must reach a U(1)$'$  Landau pole. Near the scale at which the U(1)$'$ becomes non-perturbative either the theory enters a strong coupling regime or new physical states appear which maintain the theory in a weakly coupled completion.\footnote{New physics which enters at the Landau pole does not necessarily need to take part in anomaly cancellation. The low energy theory could transition to a different weakly coupled theory which remains anomalous, and now anomaly cancelation must take place in the new theory.}  In principle such new physics could be observable at collider experiments if it occurs near the TeV scale.

The running of $g'$ is only initiated above the $Z'$ mass, thus $g'(m_Z)=g'(m_{Z'})$. At energies $Q<m_{Z'}$ running is inhibited by the $Z'$ mass, much as the Fermi constant $G_F$ does not run. Above $m_{Z'}$ the U(1)$'$ coupling strength $\alpha'\equiv g'{}^2/4\pi$ runs with the energy scale $Q$,
\beq
\frac{{\rm d} \alpha'{}^{-1}}{{\rm d} \ln Q} = -\frac{b}{2\pi}
 \hspace{10mm} {\rm with} \hspace{10mm}
 b =  \sum_f \frac{2}{3} z_f^2 +\sum_s \frac{1}{3} z_s^2~,
 \label{eqn:beta}
\eeq
where the sum runs over all U(1)$'$ charged Weyl fermions $f$ and complex scalars $s$ with charge $z_i$ that are accessible at the scale $Q$ and includes colour and representation factors. 

The scale at which $g'$ becomes non-perturbative depends not only on the field content and charge assignments, but also the masses of any new fields. Below the TeV scale, we assume that only the SM fields, and DM states $\chi_L$ and $\chi_R$ (except for Model $\sharp$1), are present. If the new fermions enter at the scale $M$, the running of $g'$ to some UV scale $\Lambda$ is described by
\beq
 \alpha'{}^{-1} (\Lambda) = \alpha'{}^{-1}(m_{Z'}) -\int^{M}_{m_{Z'}}  \frac{b_{m_Z}}{2\pi}~{\rm d}  \ln Q  - \int^{\Lambda}_{M}  \frac{b_{m_Z}+b_{M}}{2\pi}~{\rm d}  \ln Q~,
 \label{RGE2}
\eeq
where $b_{m_Z}$ and $b_{M}$ are defined as in eq.~(\ref{eqn:beta}), but now for $b_{m_Z}$ the sum is over the SM states and DM, and for $b_{M}$ we sum over only the new fermions required by anomaly cancellation. 
Specifically, the U(1)$'$ coupling runs non-perturbative ($\alpha'(\Lambda_{\not P})\sim1$) at the scale $Q=\Lambda_{\not P}$
\beq
\Lambda_{\not P}= M\exp\left[\frac{1}{b_M+ b_{m_Z}}\left( b_{m_Z} ~{\rm log} \left[\frac{m_{Z'}}{M}\right]
  +\frac{2\pi}{\alpha'(m_Z)}-\frac{2\pi}{\alpha'(\Lambda_{\not P})}\right)\right]~.
\label{RGEL}
\eeq

There could be additional vector-like pairs of fermions, or new scalars, charged under U(1)$'$ which will increase running without altering the anomaly cancellation requirements. Indeed, one typically introduces scalars charged under U(1)$'$  to give masses to the exotics through a Higgs mechanism, as discussed in Section~\ref{Sec4}. Furthermore, certain states charged under U(1) might be integrated out at some scale $\Lambda_*$ leading to higher dimension operators in the low energy theory (as may be useful to give mass to some SM fermions or exotics, cf.~eq.~\ref{uv}). If $\Lambda_*<\Lambda_{\not P}$, however, then these states must also in principle be included in the running of the U(1)$'$ gauge coupling above $\Lambda_*$. Here for simplicity we include only the new fermions required by anomaly cancellation in the U(1)$'$ gauge coupling running. Thus, our constraints may be weaker than in a complete model, but qualitatively they usually will not change. Note that for the spectra we consider, the pole for  U(1)${}_Y$ always lies above $\Lambda_{\not P}$.

\begin{figure}[t!]
\vspace{-5mm} 
\begin{center}
\includegraphics[width=0.37\textwidth]{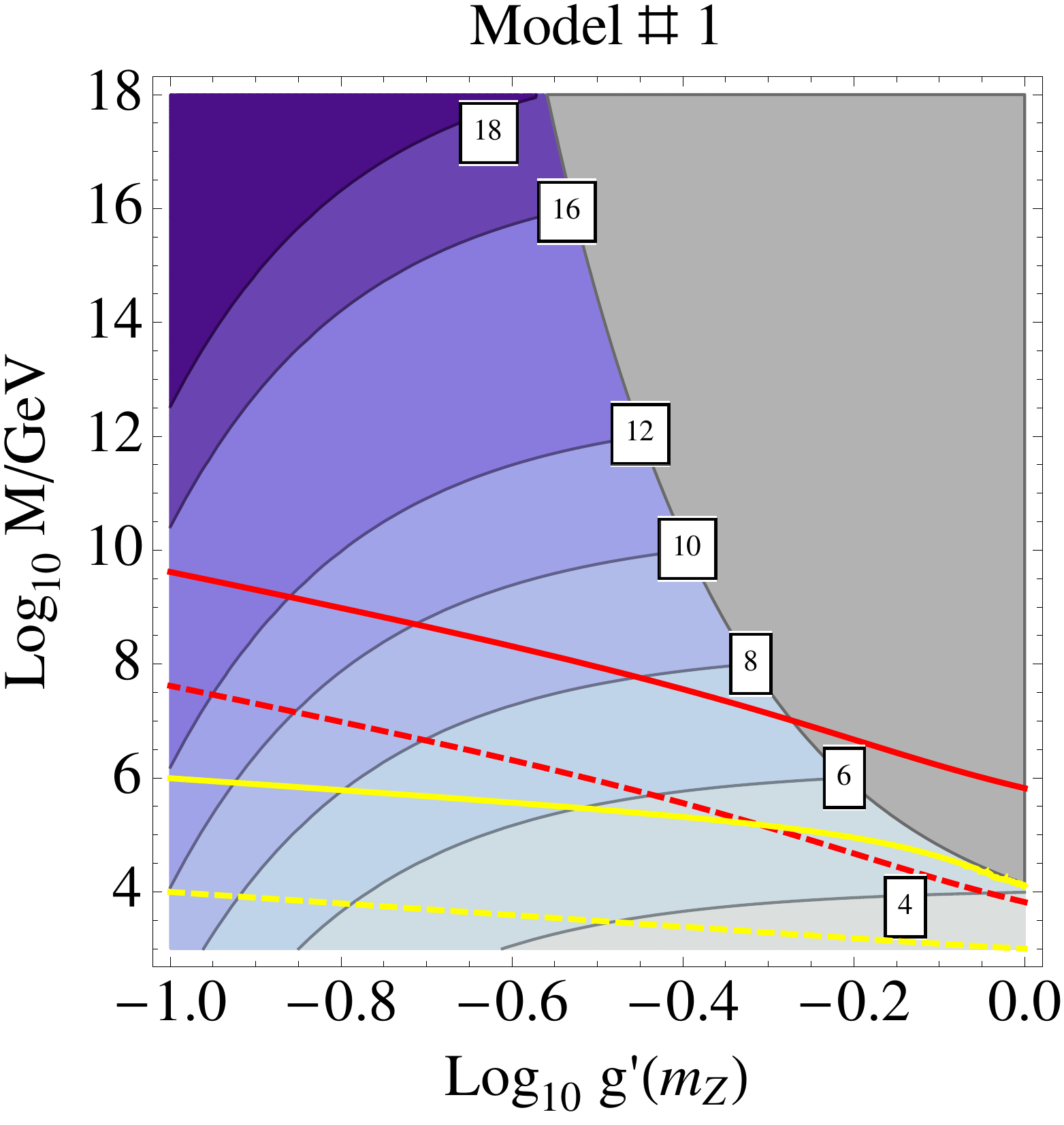}
\hspace{10mm} 
\includegraphics[width=0.37\textwidth]{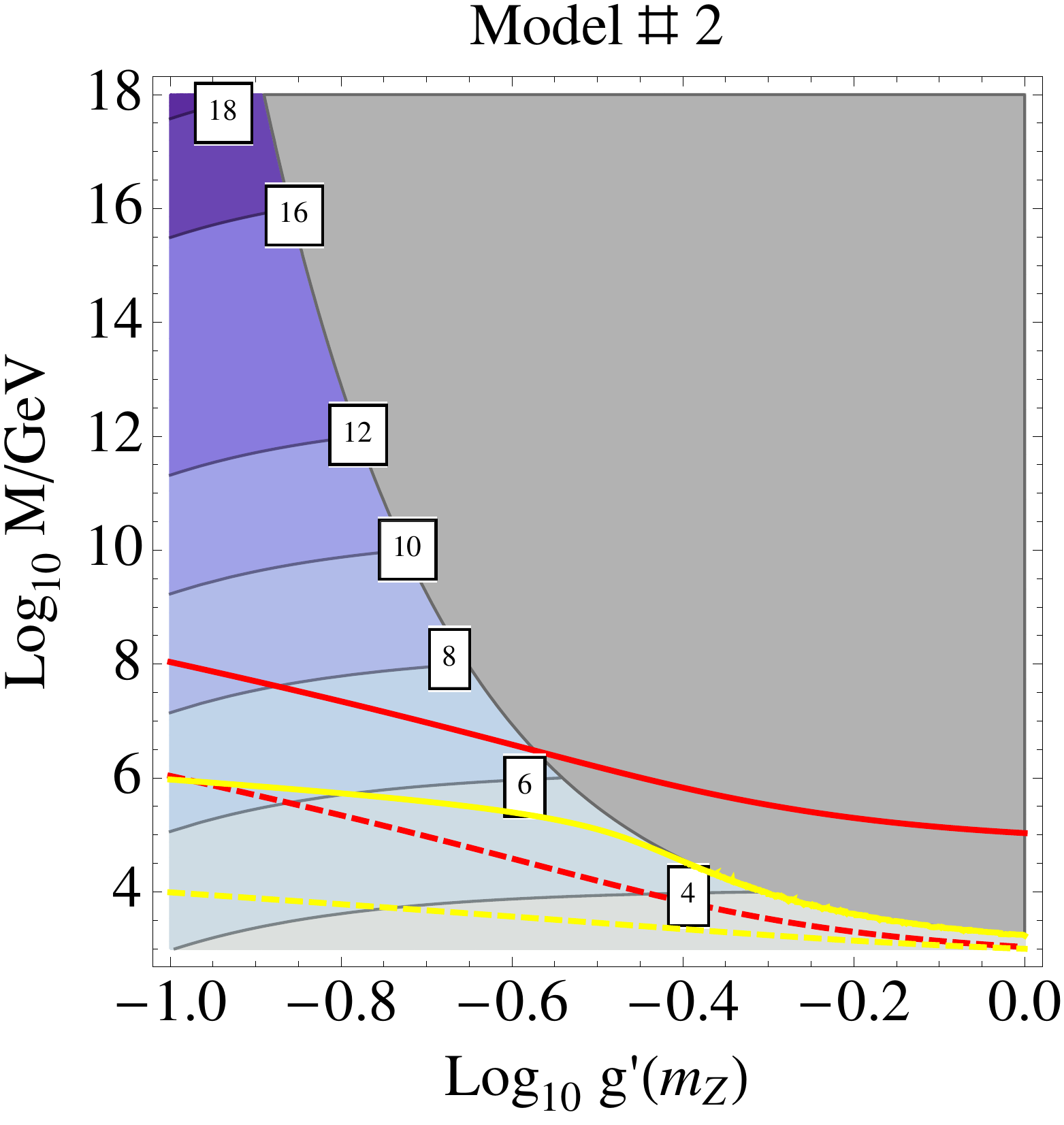}\\[5pt]
\includegraphics[width=0.37\textwidth]{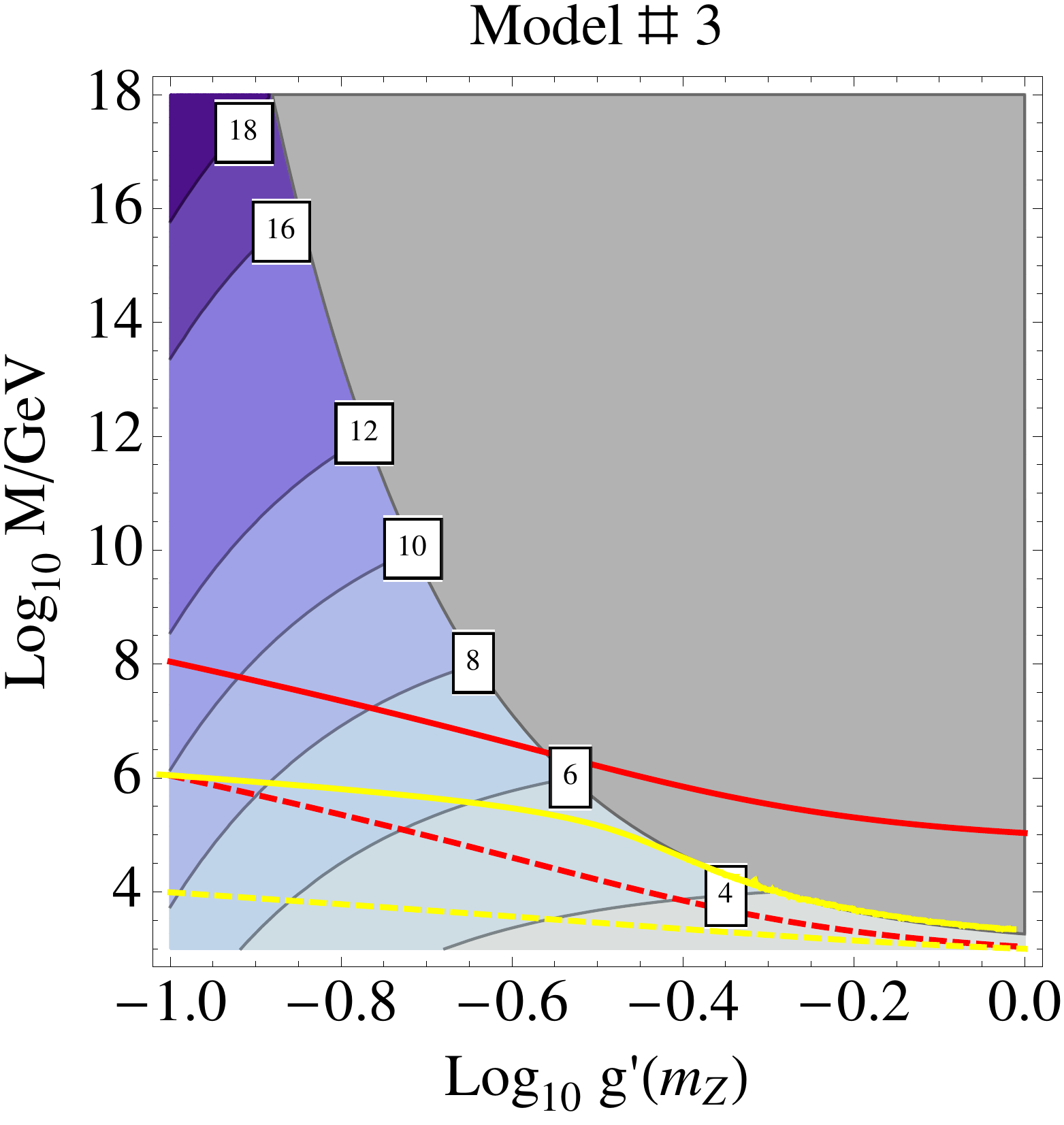}
\hspace{10mm} 
\includegraphics[width=0.37\textwidth]{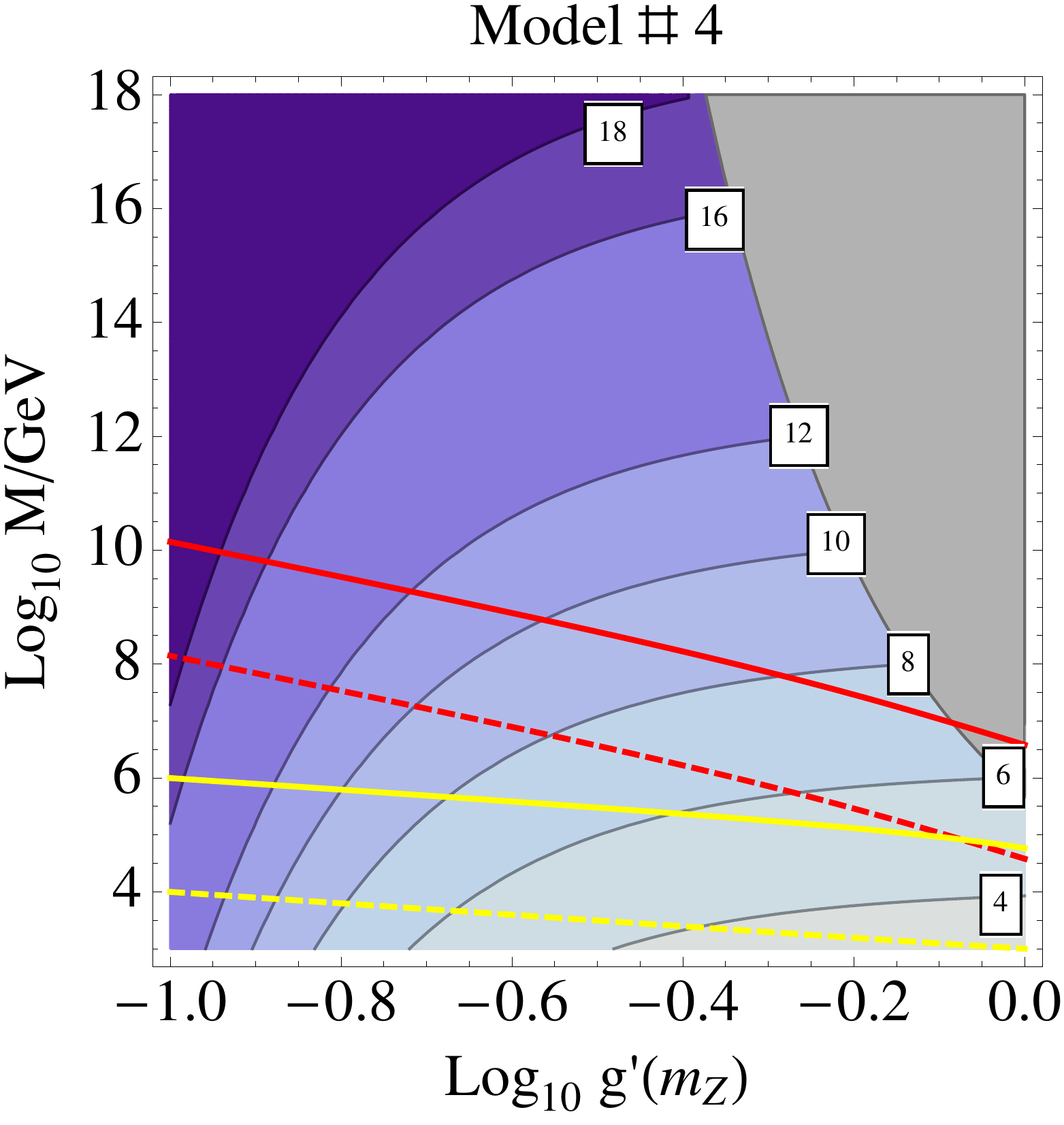}\\[5pt]
\includegraphics[width=0.37\textwidth]{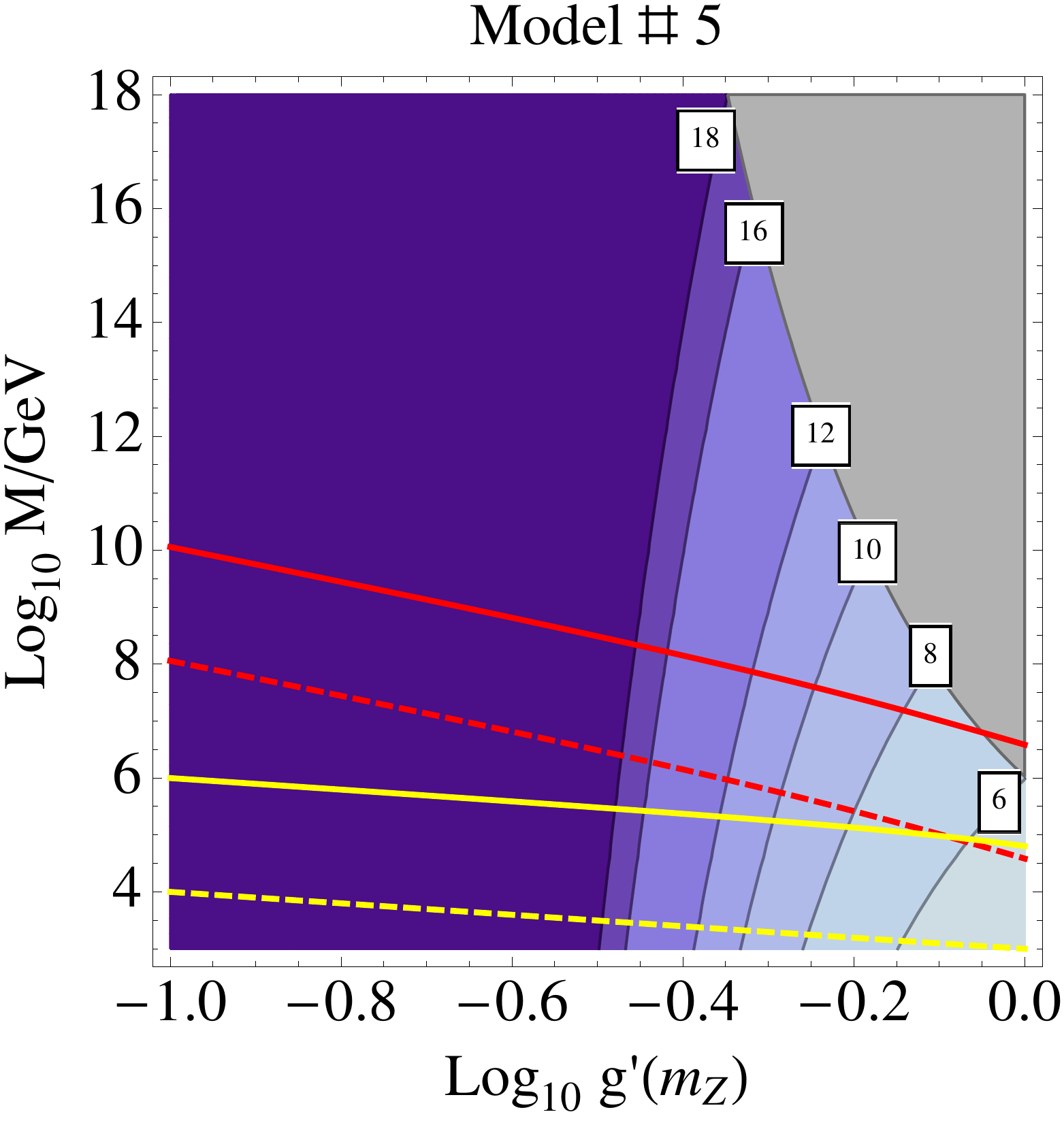}
\hspace{10mm} 
\includegraphics[width=0.37\textwidth]{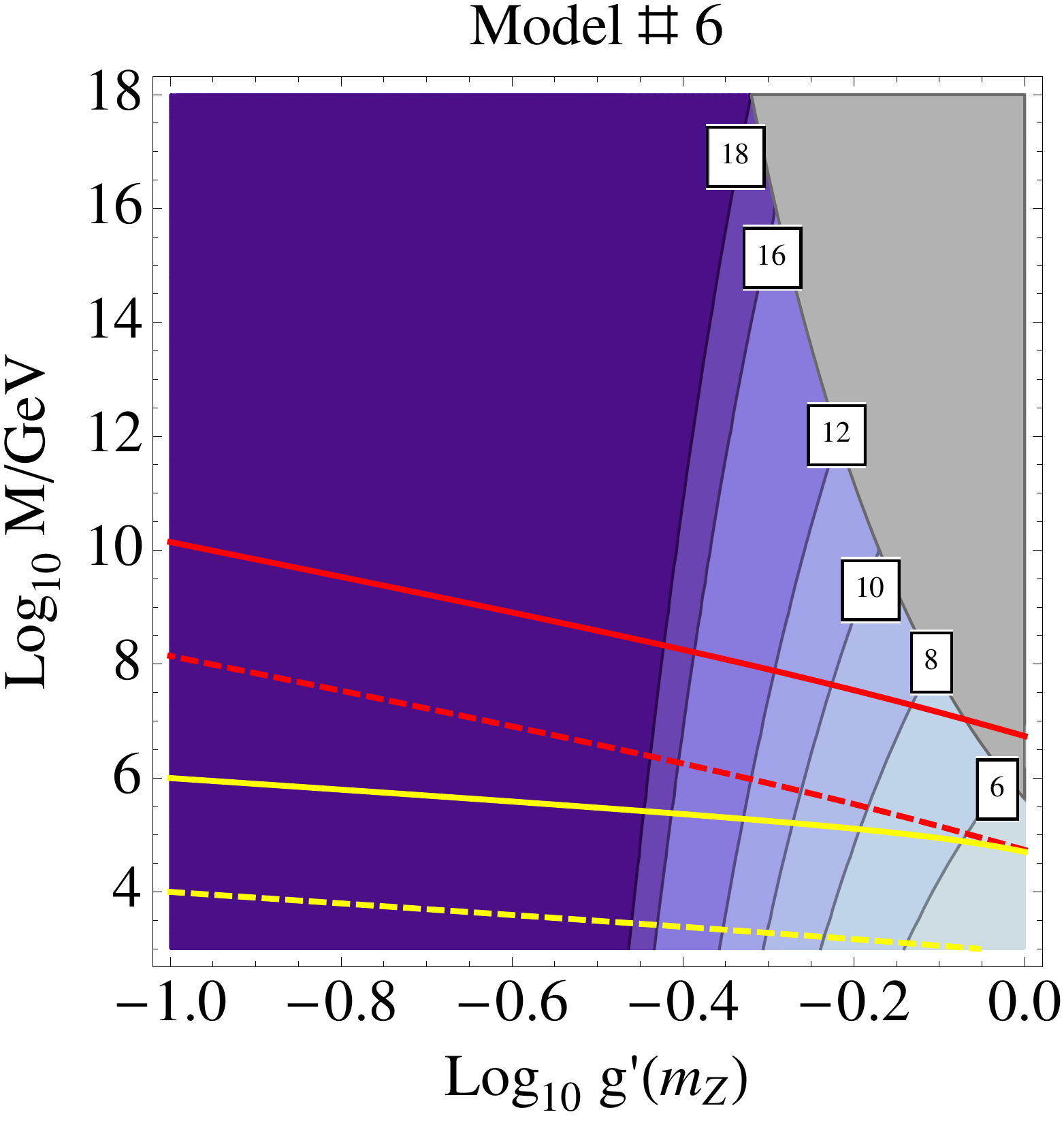}\\[5pt]
\vspace{-4mm} 
\caption{
Assuming $m_{Z'}\simeq 1$ TeV and $m_{\rm DM}\sim m_Z$ the {\sc purple contours} show the scale $\Lambda_{\not P}$ at which the U(1)$'$ coupling $g'$ runs non-perturbative in Models $\sharp$1-$\sharp$6. The boxes show values of Log${}_{10}[\Lambda_{\not P}/{\rm GeV}]$. The scale $\Lambda_{\not P}$ depends on $g'(m_Z)$, and $M$, the scale of the anomaly cancelling fermions. Only the contributions from the SM fermions, DM, and anomaly cancelling fermions are used in the RG evolution. The {\sc grey} region indicates that $\Lambda_{\not P}<M$, and new physics enters at $\Lambda_{\not P}$ regardless of $M$. The {\sc red curves} show the maximum scale at which exotics must enter to prevent the loss of renormalisability $\Lambda_{\not R}$ for $m_{Z'}\sim1$ TeV  ({\sc dashed}), and 100 TeV ({\sc solid}). For exotics which acquire mass through $v'$, the VEV that breaks U(1)$'$, $M$ and $m_Z$ are related. We show the restriction $M\lesssim v'\simeq m_{Z'}/g'$, as {\sc Yellow curves}  for $m_{Z'}\sim1$ TeV  ({\sc dashed}), and 100 TeV ({\sc solid}).
\label{fig1}}
\end{center}
\end{figure}

Figure~\ref{fig1} shows the RG evolution of the U(1)$'$ coupling $g'$ in Models $\sharp$1-$\sharp$6 with the assumption that $m_{Z'}\sim$ TeV. The {\sc purple} contours indicate the scale $\Lambda_{\not P}$ at which $g'$ becomes non-perturbative for a given $g'(m_Z)$ and $M$. If $g'$ starts out sufficiently small at the electroweak scale and the new fields are heavy, the Landau pole is reached only at very high scales.  But observe that for weak scale couplings $g'(m_Z)\sim 0.1-1$ Landau poles can be a concern for all models we consider. At the threshold of strong coupling $\Lambda_{\not P}$ one expects new physics with observable consequences.  In particular, TeV scale non-perturbativity is evident in Models $\sharp$1-4 with weak scale exotics, as indicated by the lightest contours in Figure \ref{fig1}.  On the other hand, for sufficiently small $g'(m_{Z})$ the U(1)$'$ coupling may not run to strong coupling until above $10^{18}$ GeV (which coincides with the Planck scale), as indicated by the darkest contour.  

In the {\sc grey} region, $\Lambda_{\not P} < M$ and the running due to the low energy content alone will cause $g'$ to reach its pole before the anomaly cancelling fermions enter, so that new physics is expected at this scale regardless of anomaly cancellation considerations. The boundary of this region saturates this bound, thus $M=\Lambda_{\not P}$ and $g'(M)=\sqrt{4\pi}$, and the scale at which $g'$ runs non-perturbative  can be read from the LH axis. The coupling at the weak scale is determined by the RG evolution, according to eq.~(\ref{RGEL}), from the UV scale $M=\Lambda_{\not P}$
\beq
\alpha'(m_Z)=\left(\frac{1}{\alpha'(\Lambda_{\not P})}- \frac{b_{m_Z}}{2\pi} ~{\rm log} \left[\frac{m_Z}{\Lambda_{\not P}}\right]
\right)^{-1}~.
\eeq
In order words, the trajectory of the boundary curve enveloping the grey region relates the scale $M$ and $g'(m_Z)$ by the RG evolution backward from $g'(M)=\sqrt{4\pi}$ with only SM particle content plus the added DM (if present).

\afterpage{\clearpage}

\subsection{The Non-Renormalisable Limit}

If a set of fermions is anomalous at a given energy scale, it should be anticipated that this is an EFT and at some higher scale $M$ additional fermions (or another mechanism) enters to cancel the anomalies. Below the scale $M$ the heavy chiral fermions which are integrated out generate Wess-Zumino terms which cancel the apparent anomalies in low energy theory \cite{Sterling:1981za,D'Hoker:1984ph,Preskill:1990fr}.  However, the cutoff of the EFT in which the anomaly cancelling fermions are integrated out can not be made arbitrarily high without losing calculability. If a gauge anomaly remains uncancelled it eventually results in a loss of renormalisability. 
For an EFT with gauge anomalies there is a fundamental cutoff $\Lambda_{\not R}$ at which renormalisibility is lost, and for an anomalous U(1) gauge theory this is given by  \cite{Preskill:1990fr}
\beq
M\lesssim m_{Z'}\left(\frac{64\pi^3}{|g_{\not R}'{}^3\mathcal{A}_{Z'Z'Z'}|}\right)\equiv\Lambda_{\not R}~,
 \label{notR}
\eeq
where  $g_{\not R}'\equiv g'(\Lambda_{\not R})$ and $\mathcal{A}_{Z'Z'Z'}={\rm Tr}[z^3]$ is the U(1)$'{}^3$  anomaly coefficient in the EFT below the scale of the exotics $M$.  Therefore, the requirement that the gauge theory remains renormalisable places an upper limit on the scale of the anomaly cancelling exotics $M$. 

The anomaly cancelling exotics must enter at, or prior to, the scale $\Lambda_{\not R}$, as determined by eq.~(\ref{notR}). Moreover, if we suppose that the exotics enter at the highest possible scale, $M=\Lambda_{\not R}$, then eq.~(\ref{notR}) implies the following model independent upper bound on the coupling for a given set of charges
\beq
\frac{|g'_{\not R}{}^3\mathcal{A}_{Z'Z'Z'}|}{64\pi^3}\lesssim \frac{m_{Z'}}{M}\lesssim1~,
\label{ine}
\eeq 
since a reliable EFT must satisfy $m_{Z'}\lesssim\Lambda_{\not R}$. For $m_{Z'}\sim M$ this requirement does not significantly constrain the parameter space, but stronger bounds are obtained for specific values of $m_{Z'}$. To derive a useful constraint we take a range  of values for $M$ and find the coupling $g_{\not R}$ which saturates the inequality (\ref{ine}) for $m_Z'=$1 TeV and 100 TeV. Running $g_{\not R}'$ from $M$ to the scale $m_{Z'}$ (via eq.~(\ref{RGE2})), gives a bound on the low energy coupling $g'(m_Z)$. 

For $m_{Z'}\ll M$ the exotics must enter to prevent the loss of renormalisability prior to the scale at which one anticipates a Landau pole, i.e.~$\Lambda_{\not R}\ll \Lambda_{\not P}$.   Furthermore, the mass scale of the exotics is characteristically  set by the U(1)$'$ breaking scale $v'$, i.e.~$M\lesssim v'\simeq \frac{m_{Z'}}{g'(m_{Z'})}$ (cf.~eq.~(\ref{limv})). In this case the exotics must typically enter earlier than dictated by perturbativity or renormalisibility considerations. The scale of EFT breakdown $\Lambda_{\not R}$ and the requirement that $M\lesssim v'$ are both shown in Figure~\ref{fig1} for $m_{Z'}=1$ TeV,  and 100 TeV.


\section{Dark Matter Freeze-out via an Axial Vector}
\label{Sec6}

One of the leading motivations for considering a new abelian gauge boson with only axial vector couplings to the SM fermions is the prospect of providing a potential mediator between DM and SM fermion interaction. As such it is of interest to consider the possibility of successful thermal freeze-out of the DM, with the relic density of DM determined by annihilation to SM states mediated by the axial vector. Here we will restrict ourselves to the scenario in which the charges of the DM $\chi$ and SM fermions $f$ are fixed to be Model $\sharp2$ of Table \ref{Tab5}. Further, we assume that the Higgs is not charged under U(1)$'$, and only consider $\bar\chi \chi\rightarrow \bar f f$ annihilation.  A similar analysis could be carried out for alternative models.

\begin{figure}[t!]
\begin{center}
\includegraphics[width=0.48\textwidth]{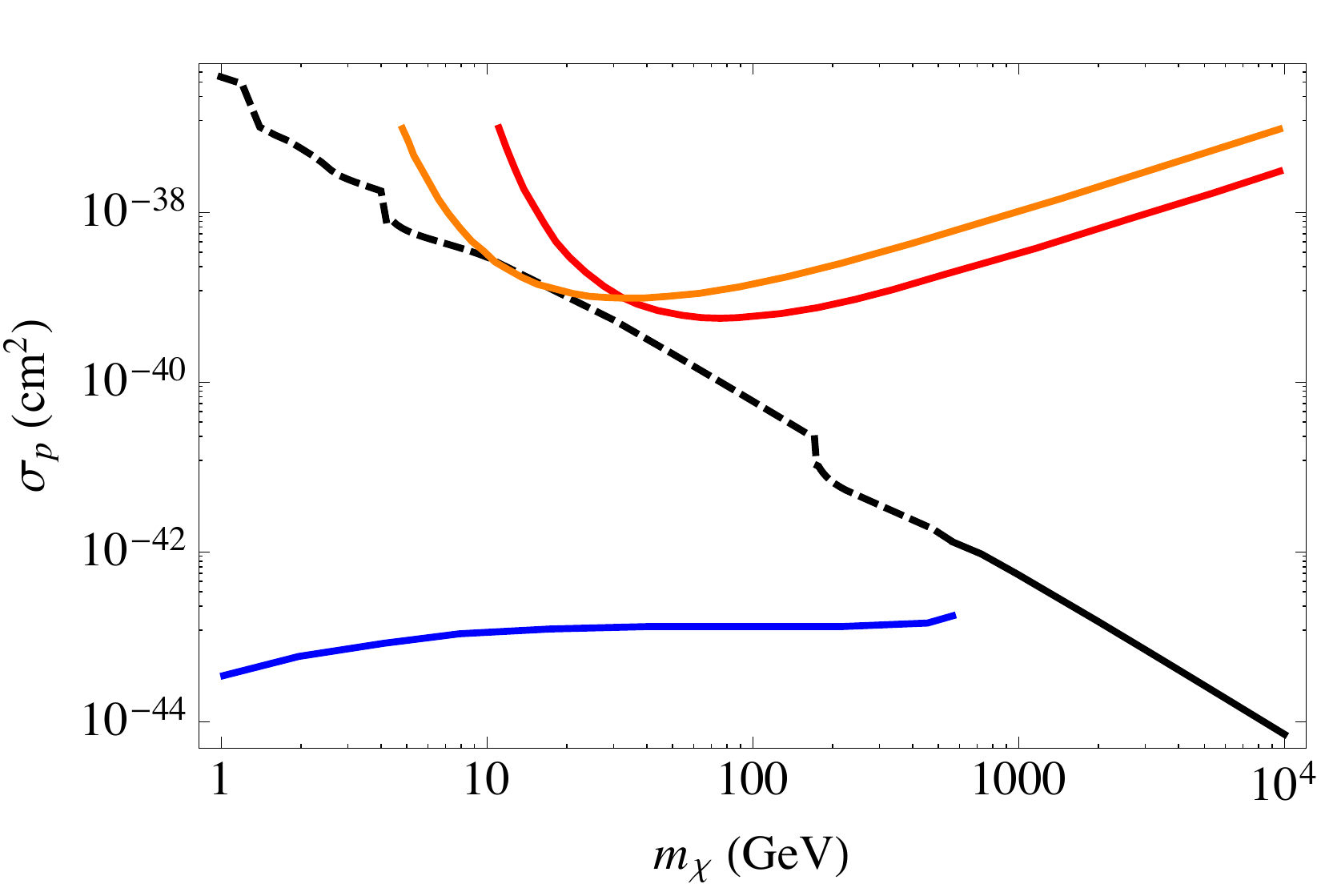}~
\includegraphics[width=0.48\textwidth]{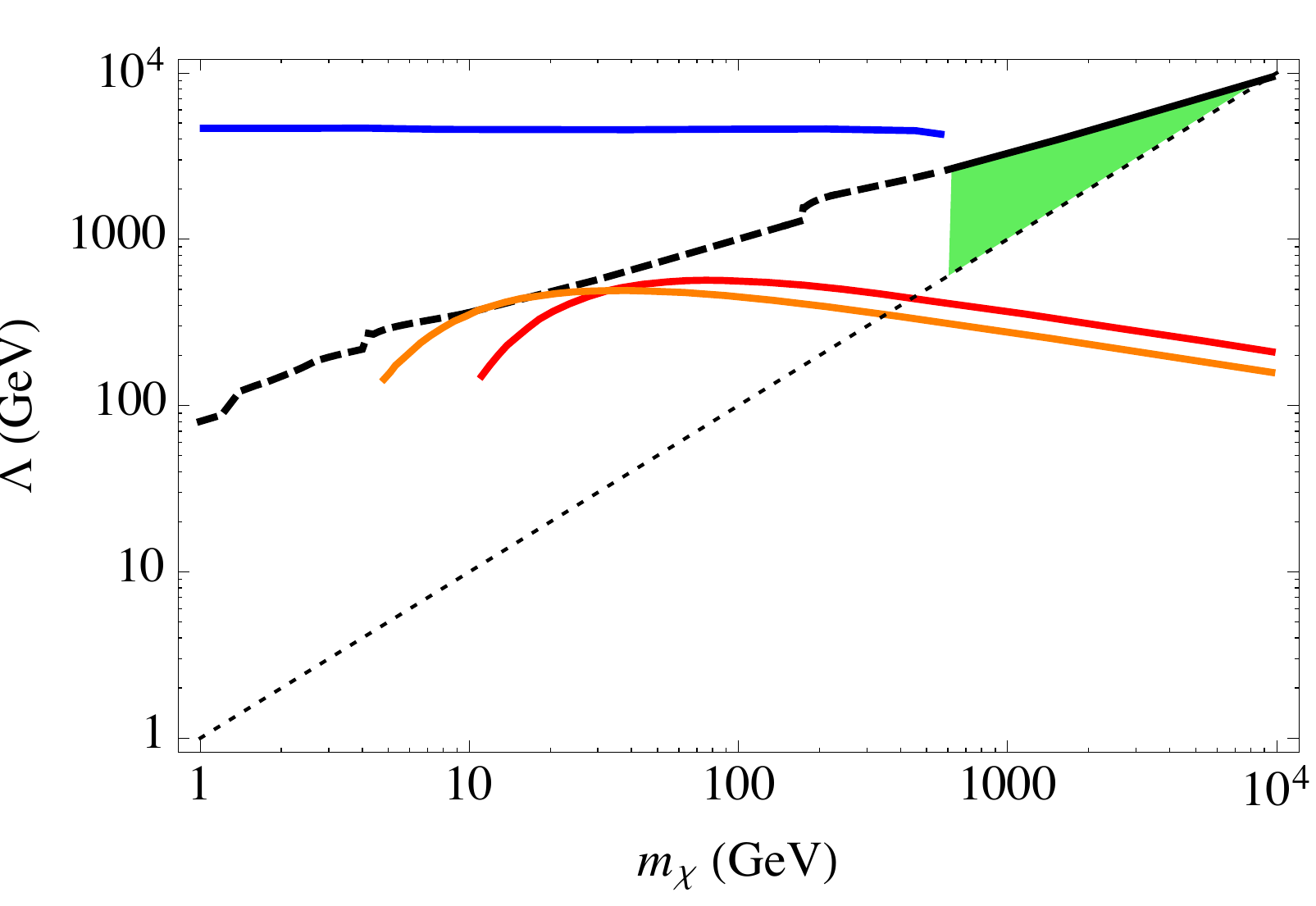}
\vspace{-4mm}
\caption{
Constraints on Dirac DM $\chi$ coupling to the SM via $\frac{1}{\Lambda^2}\bar \chi \gamma^{\mu}\gamma^5 \chi \bar f \gamma_{\mu} \gamma^5 f$ coming from {\sc pico60} ({\sc red}) \cite{Amole:2015pla}, {\sc pico2L} ({\sc orange}) \cite{Amole:2015lsj}, and {\sc CMS} monojet searches ({\sc blue}) \cite{CMS:2016pod}. The LH panel shows the $\sigma_{\rm SD}$-$m_\chi$ plane. The RH plot gives the same information in the $\Lambda$-$m_\chi$ plane. The {\sc black} curve shows the $\Lambda$ such that DM freeze-out reproduces the correct relic density, the curve is dashed where such couplings are in tension with direct searches. For DM with mass 600 GeV$\lesssim m_\chi\lesssim$10 TeV the correct relic density can be reproduced without conflicting with direct constraints. The {\sc green shaded} region indicates parameter regions where the DM density set by freeze-out is below the observed relic density.  The thin {\sc dotted line} on RH plot indicates the regime $m_{\chi}\gtrsim\Lambda $ where the EFT is not reliable.
\label{fig2}}
\vspace{-8mm}
\end{center}
\end{figure}

If the $Z'$ is heavy relative to the DM and SM states, the mediator can be integrated out yielding a dimension six operator $\frac{1}{\Lambda^2}\bar \chi \gamma^{\mu}\gamma^5 \chi \bar f \gamma_{\mu} \gamma^5 f$ connecting DM with SM fermions, with
$\Lambda\equiv m_{Z'}/g'\sqrt{(2z_q)(2z_{\rm DM})}$.
For Model $\sharp$2 we have $\sqrt{4z_qz_{\rm DM}}=6$.
The cross section for Dirac DM annihilating to SM quarks via this operator is 
\cite{MarchRussell:2012hi}
\beq
  \sigma_{\mathrm{ann}} v & =
 \frac{3m_{\chi}^2}{2\pi\Lambda^4}  \sum_q \left(1-\frac{m_q^2}{m_{\chi}^2}\right)^{1/2}
 \Bigg[ \frac{m_q^2}{m_{\chi}^2}  + v^2
 \left(\frac{8m_{\chi}^4-22m_q^2m_{\chi}^2+17m_q^4}{24m_{\chi}^2(m_{\chi}^2-m_q^2)}\right)\Bigg]+\mathcal{O}(v^4).
\eeq
where $v$ is the DM relative velocity.
 Thus the requirement that the annihilation cross section is appropriate to give the observed DM relic density constrains the magnitude of $\Lambda$ for a given DM mass $m_\chi$. Following  \cite{Kurylov:2003ra,Beltran:2008xg,Fitzpatrick:2010em}, in Figure \ref{fig2} we show the value of $\Lambda$ required to obtain the observed relic density as $m_\chi$ varies.   Note that the EFT is no longer reliable if the DM mass exceeds the cutoff, so we require $m_{\rm DM}\lesssim \Lambda$, as indicated by the dashed line in the RH panel of Figure~\ref{fig2}. This EFT requirement can be re-expressed as a constraint on the $Z'$ mass and coupling $g'$, as in eq.~(\ref{EFT}).

The operator induces spin-dependent DM-nucleon scattering and thus can be searched for via direct detection experiments. For Dirac fermion DM scattering with quarks, mediated via a heavy axial vector the spin-dependent scattering cross section with protons is  \cite{Kurylov:2003ra, Beltran:2008xg,Fitzpatrick:2010em} 
\beq 
\sigma_{p}  \approx \frac{4}{\pi\Lambda^4}\mu_p^2\left(\sum_{q=u,d,s}    \Delta^p_q \right)^2 ~,
\eeq
  where $\mu_p\equiv\frac{m_\chi m_p}{m_\chi+m_p}$ is the reduced mass, and $\Delta^{p}_q$ is the spin content of the nucleon \cite{Cheng:2012qr}; we use $\sum_q    \Delta^p_q \approx0.37$. A smaller $\Lambda$, from larger couplings or equivalently lighter mediators, corresponds to a higher scattering rate.  Given the scattering cross section $\sigma_{p}$ we can apply the current direct detection limits to the EFT, and derive a lower bound on $\Lambda$. We use the limit from the {\sc pico} experiment \cite{Amole:2015pla,Amole:2015lsj} to put a constraint on $\Lambda$, see Figure \ref{fig2}. We have not included RG running, see e.g.~\cite{DEramo:2016atc}, but this is expected to have only a mild effect in this case.
  
Additionally, indirect detection signals due to DM annihilation producing photons and neutrinos are searched for by Fermi~\cite{Ackermann:2015zua}, IceCube~\cite{Aartsen:2016exj} and Super-Kamiokande~\cite{Choi:2015ara}; these can provide complementary constraints. A full analysis is beyond the scope of this paper, however see e.g.~\cite{Beltran:2008xg,Jacques:2016dqz} for further discussion.

The axial vector can also mediate DM production through collisions of SM states, and thus searches at colliders for events with missing energy constrain the production cross section. Figure \ref{fig2} displays limits from  {\sc CMS} searches with $\sqrt{s}=13$ TeV and 12.9 fb${}^{-1}$ \cite{CMS:2016pod}. The {\sc CMS} analysis shown assumes a simplified model with $z_q g'=z_\chi g'=0.05$. In this limit the mediator is heavy enough that it is not kinematically accessible, and we cutoff the limit before on-shell effects affect the line shape; small variations in the couplings can be absorbed into $m_{Z'}$ with little impact. An EFT should give a similar limit, and in the RH plot we refashion the {\sc CMS} limit in terms of an EFT by identifying $\Lambda\equiv m_{Z'}/g'\sqrt{(2z_q)(2z_{\rm DM})}$; this is strictly only reliable for $\sqrt{s}< m_{Z'}$.
If the axial vector mass is comparable to LHC energies ($m_{Z'}\lesssim\sqrt{s}=$ 13 TeV)  the EFT may break down and this requires a UV completion, examples of which we have outlined above. For discussions of on-shell medaitor effects see e.g.~\cite{MarchRussell:2012hi,Buchmueller:2014yoa,deSimone:2014pda,Lebedev:2014bba,Arcadi:2014lta,Chala:2015ama,Kahlhoefer:2015bea}.

For 600 GeV$\lesssim m_\chi\lesssim$10 TeV the DM relic density can be reproduced without conflict with constraints.  This viable parameter space corresponds to 1 TeV$\lesssim\Lambda\lesssim10$ TeV, thus for moderate couplings (say $0.01\lesssim g'\lesssim1$)  the axial vector is of order 60 GeV $\lesssim m_{Z'}\lesssim60$ TeV. However, LHC constraints typically require $m_{Z'}\gtrsim1$ TeV for couplings $g'\sim\mathcal{O}(0.1)$ \cite{ATLAS-CONF-2016-069}.  

The above discussion  assumes the DM relic abundance is set by freeze-out, in alternative scenarios these requirements will vary. For instance, in Asymmetric Dark Matter \cite{Zurek:2013wia} one desires that the density of DM-antiDM pairs is reduced below the observed relic density, such that a DM-antiDM asymmetry can be responsible for the late time abundance. Thus this scenario requires even more efficient annihilation, which shrinks the viable parameter space; see  \cite{MarchRussell:2012hi,Buckley:2011kk} for studies of DM annihilation via $\bar \chi \gamma^{\mu}\gamma^5 \chi \bar f \gamma_{\mu} \gamma^5 f$ in Asymmetric Dark Matter.
Furthermore, in parameter regions in which the DM density is not reduced below the observed relic density, the correct abundance might still be obtained via other mechanisms, e.g.~entropy injection (e.g.~\cite{Gelmini:2006pw}), DM freeze-in (e.g.~\cite{Hall:2009bx,Chu:2011be,Elahi:2014fsa}), or thermal inflation (e.g.~\cite{Hui:1998dc,Lyth:1995ka}).


\section{Discussion}
\label{Sec7}

Axial vectors have been motivated in a number of different contexts. For instance, they appear commonly as mediators for DM interactions with SM states. Whilst many studies consider scenarios with axial vector gauge bosons, they often neglect to confront the challenges of anomaly cancellation. Ensuring that a model is anomaly free is crucial for the gauge theory to be consistent, and successful anomaly cancellation typically requires new states which are charged under the SM gauge group. Moreover, as we have argued here, these new fermions can not be arbitrarily separated in mass from the axial vector.  

Thus it is important to consider UV completions as these new exotics required for anomaly cancellation are potentially observable at colliders.  In particular, unless U(1)$'$ charges differ in each SM generation, an axial vector which couples to quarks requires new coloured fermions for anomaly cancellation. In the case of a universal axial vector with couplings to DM that thermally produce the observed relic density, the new coloured fermions should be at the 1-10 TeV scale, and can be probed in the future.

Additionally, when the $Z'$ is accessible at colliders, limits arise from resonance searches. Current LHC limits from dijet (dilepton) searches for axial vectors with $g'\sim0.1$ typically require $m_{Z'}\gtrsim1$ TeV ($m_{Z'}\gtrsim3$ TeV) \cite{ATLAS-CONF-2016-069,Chala:2015ama,Jacques:2016dqz,Khachatryan:2014fba}, which would weaken somewhat if the $Z'$ has a large branching fraction to DM, or not be applicable if the $Z'$ is leptophilic (leptophic). Both of these scenarios occur in the models we have discussed. In the case of $Z$-$Z'$ mixing, there are also limits from electroweak precision constraints \cite{Erler:2009jh,Babu:1997st,Agashe:2014kda}. Moreover, if there are exotic Higgs states to give mass to the new fermions, this can lead to other bounds such as variations in Higgs couplings to SM states, see e.g.~\cite{Curtin:2013fra}, or contributions to the invisible Higgs width \cite{Shrock:1982kd,CMS:2016rfr}.  Furthermore, after U(1)$'$ breaking states with the same SM quantum numbers will generically mix (and beforehand if the states have identical charges) this $f$-$f'$ mixing is constrained by electroweak precision and flavour observables. However, these constraints are typically model dependent, see e.g.~\cite{Atre:2011ae}. 
A full analysis of the constraints, and model dependence, of each of the scenario considered here is beyond the scope of this work, but in a forthcoming paper we will examine some of these phenomenological issues for the $t$-$b$-philic case (Model $\sharp$6 of Table \ref{Tab2} and related scenarios). 

In conclusion the purpose of this paper has been two-fold: Firstly we have provided anomaly free, UV complete reference models for axial vector gauge bosons coupling to SM fermions. In the course of deriving the anomaly-free sets of fermions we have explored a number of general methods for constructing such models. Secondly, we wished to highlight that in neglecting the additional states required for anomaly cancellation, one omits a number of potentially important constraints, such as collider searches for anomaly cancelling exotics, the need for new scalars to give mass to exotics, the possibility of low U(1)$'$ Landau poles, and potentially the loss of renormalisibility, all which should be taken into consideration in any full model. 

\vspace{30mm}
\section*{\bf Acknowledgements~}
We thank B.~Dobrescu, M.~Goodsell,  A.~Hook, A.~Katz, J.~March-Russell, and S.~West for useful interactions. This work is partially supported by the U.S.~Department of Energy under contracts DE-AC02-06CH11357, DE-FG-02-12ER41811, and DE-SC0015634. AI is grateful to the Aspen Center for Physics, which is supported by National Science Foundation grant PHY-1066293, for its hospitality during the completion of this work. JU is grateful to the LPTHE, Paris VI, and the CERN Theory Division for their hospitality.

\vspace{3mm}

\appendix

\newpage
\section{A Selection of Anomaly Free Vector Models}
\label{ApC}

We provide Table \ref{Tab6} of the charge assignments which lead to anomaly cancellation for the case of pure vector couplings to the SM fermions (and dark matter), for analogues of Models $\sharp$1-$\sharp$6. This is given both for completeness and to demonstrate that the axial vector case typically requires far more exotics in order  to arrange for anomaly cancellation compared to the vector case. This also highlights that there is no need of coloured exotics in the pure vector case. 

\vspace{8mm}

\begin{table}[h!]
\begin{center}
\def\str{\vrule height8pt width0pt depth5pt}
\begin{tabular}{| c | c | c | c | c  | c | c|}
    \hline\str
    ~~~Field ~~~&~~~ $\sharp1V$ ~~~&~~~ $\sharp2V$~~~&~~~ $\sharp3V$~~~ &~~~$\sharp4V$ ~~~ &~~~ $\sharp5V$~~~&~~~ $\sharp6V$~~~ \\[2pt]
     \hline\str
    $z[Q_L]$          & 1       &  1      & 1      &  0     &     1   & 1      \\
    $z[u_R]$          &  1      &  1     &  1      &  0     &     1   &   1     \\
    $z[d_R]$          & 1     &  1       & 1     &  0      &     1   &     1     \\
    $z[L_L]$          &  1     &   1      & 0      &  1      &    1    &    0     \\
    $z[e_R]$          &  1     &  1     &0      & 1         &    1    &     0   \\
    $z[{\chi}_{L}]$   &    -      &  1    &  1    &  1       &   1     &     1      \\
    $z[{\chi}_{R}]$   &    -    &  -1  & -1    & -1        &    -1    &     -1    \\
            \hline\str
    $z[L'_{L}]$         &   -1   & -6    & -4       & -1       &  -2      &   -1          \\
    $z[L'_{R}]$         &   11   & 6    & 5     &  2        &    2    &      2      \\
    $z[e'_{L}]$          &  11   &  6    & 5    &  2        &     2   &      2   \\
    $z[e'_{R}]$         &   -1   &  -6   & -4     & -1          &  -2      &  -1      \\
        \hline\str
    $z[\nu_R^{(1)}]$   &  1   &  -1  & 1   &  1          &    1    &      1    \\
    $z[\nu_R^{(2)}]$   &  2   &  2   &  -4   &  -2         &  -2      &    -2        \\
    $z[\nu_R^{(3)}]$   &  -3  & -5   &  -5   &  -         &   -     &    -        \\
    $z[\nu_R^{(4)}]$   &  -7  & -7  &   -   &  -          &    -    &     -   \\
    $z[\nu_R^{(5)}]$   &  -10 &  -   &  -    &  -       &     -   &       -  \\
         \hline\str
    $N[\nu_R^{(1)}]$   &   1  & 7  & 2  & 4          &    3    &    1        \\
    $N[\nu_R^{(2)}]$   &   5  & 6 & 1   & 1        &   2     &    1     \\
    $N[\nu_R^{(3)}]$   &   1  & 1 & 1   & -        &   -     &     -   \\
    $N[\nu_R^{(4)}]$   &   1 &  1  & -  & -         &   -     &      -   \\
    $N[\nu_R^{(5)}]$   &   1 & -  & -    & -       &    -    &       -  \\
     \hline
\end{tabular}
\renewcommand{\thetable}{A}
\caption{Similar to Table \ref{Tab5}, charge assignments for Models $\sharp$1-$\sharp$6 but for the case of a gauge boson with pure vector couplings to states (as can be seen from the charge assignments).} 
\label{Tab6}
\end{center}
\end{table}
\afterpage{\clearpage}


\newpage

\section{An Alternative Set of Anomaly Free Axial Vector Models}
\label{ApB}

\vspace{3mm}

In this appendix we give alternative anomaly free sets of fermions for the case in which the SM fermions (and dark matter) have only axial vector coupling with a new U(1)$'$ gauge boson. These charges assignments are derived using the method of \cite{Batra:2005rh}, see Section~\ref{s32}. Whilst some coloured exotics are removed, the price is the introduction of a multitude of RH neutrinos:

\vspace{10mm}


\begin{table}[h!]
\begin{center}
\def\str{\vrule height12pt width0pt depth5pt}
\begin{tabular}{| c | c | c | c | c  | c | c| c | c|}
    \hline\str
    ~~~Field ~~~&~~~ $\sharp1b$ ~~~&~~~ $\sharp2b$~~~&~~~ $\sharp3b$~~~ &~~~$\sharp4b$ ~~~ &~~~ $\sharp5b$~~~&~~~ $\sharp6b$~~~ \\[2pt]
     \hline\str
    $z[Q_L]$          & 1       &  1      & 1      &  0   &   1   &   1  \\ [1.5pt]
    $z[u_R]$          &  -1      &  -1     & -1      &  0  &   -1   &  -1     \\ [1.5pt]
    $z[d_R]$          & -1     &  -1       & -1     &  0    &   -1   &  -1     \\[1.5pt]
    $z[L_L]$          &  1     &    1      & 0      &  1    &   1   &   0   \\[1.5pt]
    $z[e_R]$          &  -1     &  -1     &0      & -1     &   -1   &   0    \\[1.5pt]
    $z[{\chi}_{L}]$   &    -      &  1    &  1    &  1     &  1    &   1     \\[1.5pt]
    $z[{\chi}_{R}]$   &    -      &  -1  & -1    & -1    &   -1   &   -1     \\[1.5pt]
            \hline\str
    $z[d'_{L}]$         &   -6   &-6     & -6     &  -   &   -2   &  -2      \\[1.5pt]
    $z[d'_{R}]$        &    6   & 6     & 6       &  -   &    2  &    2    \\[1.5pt]
    $z[L'_{L}]$         &  -6   & -6    & 1       & 1     &  -2    &   -1       \\[1.5pt]
    $z[L'_{R}]$         &  6    & 6    & 10     &  4   &  2    &    2        \\[1.5pt]
    $z[e'_{L}]$          &  -    &  -     & 18    & -4   &    -  &     2    \\[1.5pt]
    $z[e'_{R}]$         &  -     & -     & 15     & -1   &   -   &     1     \\[1.5pt]
        \hline\str
    $z[\nu_R^{(1)}]$   &  2   &  2   & 1   &  -1      &    1  &      1   \\[1.5pt]
    $z[\nu_R^{(2)}]$   &  -3  &  -3 &  -2   &  2      &   -4   &      2    \\[1.5pt]
    $z[\nu_R^{(3)}]$   &  -5  &   -5 &  -4   &  -6    & -     &      -3       \\[1.5pt]
    $z[\nu_R^{(4)}]$   &  -10& -10 &  -9    &  -     &  -    &    -    \\[1.5pt]
    $z[\nu_R^{(5)}]$   &  -    &  1   &  -    &  -   &   -   &    -    \\[1.5pt]
         \hline\str
    $N[\nu_R^{(1)}]$   &   8  & 8  & 2  & 2      & 1     &    1       \\[1.5pt]
    $N[\nu_R^{(2)}]$   &   2  & 2 & 1   & 5    &    1  &     1   \\[1.5pt]
    $N[\nu_R^{(3)}]$   &   1  & 1 & 1   & 1    &   -   &    2   \\[1.5pt]
    $N[\nu_R^{(4)}]$   &   2 & 2  &  1  & -    &    -  &     -    \\[1.5pt]
    $N[\nu_R^{(5)}]$   &   -  & 2  & -    & -   &  -    &    -    \\[1.5pt]
     \hline
\end{tabular}
\renewcommand{\thetable}{B}
\caption{Similar to Table \ref{Tab5}, alternative charge assignments  $z[f]$ and multiplicities $N[f]$ of states which give anomaly free spectra for Models $\sharp$1-$\sharp$6 derived using the method of \cite{Batra:2005rh}.}
\label{Tab7}
\end{center}
\end{table}
\afterpage{\clearpage}



\newpage

\section{Algebraic Construction of Axial Vector Examples}
\label{ApA}

Next we give a worked example of the algebraic construction of  \cite{Batra:2005rh} discussed in Section~\ref{s32} and \ref{ApB}.  We consider the scenario in which there are $n_G$ generations charged under U(1)$'$, each generation with identical charge assignments, such that the SM fermions have axial couplings to the new vector boson, and we  include chiral fermion DM states which also couple axially to the U(1)$'$ gauge boson. Following  \cite{Batra:2005rh}, the requirement that  the anomaly cancellation occurs for  U(1)${}_Y^2\times$U(1)$'$, SU(2)${}^2\times$U(1)$'$, and SU(3)${}^2\times$U(1)$'$ implies
\begin{equation*}
(z_{d'_L}-z_{d'_R})=-4n_G z_q,
\hspace{8mm}
(z_{L'_L}-z_{L'_R})=-n_G (z_l + 3 z_q),
\hspace{8mm}
(z_{e'_L}-z_{e'_R})=n_G (z_q - z_l)~. 
\end{equation*}
Taking the sum of the LH and RH charges to be a linear combination of the U(1)$'$ charges of the SM fields
$(z_{X'_L}+z_{X'_R})=C^X_{1}z_q+C^X_{2}z_{\rm DM}+C^X_{3}z_l$, for $X=d,L,e$ and where the $C^X_i$ are arbitrary integers, it follows that the charges of the exotics can be expressed in terms of the  U(1)$'$ charges of the SM fields
\beq
z_{d'_L}&=
\frac{1}{2} \Big[-C^L_2 z_{\rm DM} + C^d_3 (z_l - 9 z_q) - 2 C^e_1  z_q - 4 n_G z_q
\Big]
\\
z_{d'_R}&=
\frac{1}{2} \Big[-C^L_2 z_{\rm DM} + C^d_3 (z_l - 9 z_q) - 2 C^e_1 z_q + 4 n_G z_q
\Big]
   \\
z_{L'_L}&=
\frac{1}{2} \Big[C^L_2 z_{\rm DM} +  C^e_1 (3 z_q -  z_l ) - 4 C^d_3 (z_l - 3 z_q)  -  n_G (3 z_q  + z_l )
\Big]
  \\
z_{L'_R}&=
\frac{1}{2} \Big[C^L_2 z_{\rm DM} +  C^e_1 (3 z_q -   z_l ) - 4 C^d_3 (z_l - 3 z_q)  +  n_G (3 z_q +  z_l)
\Big] \\
z_{e'_L}&=
\frac{1}{2} \Big[-C^L_2 z_{\rm DM} + 4 C^d_3 z_l + C^e_1 (z_q  +  z_l ) - n_G (z_l - z_q)
\Big] \\
z_{e'_R}&=\frac{1}{2} \Big[-C^L_2 z_{\rm DM} + 4 C^d_3 z_l + C^e_1 ( z_q + z_l) + n_G( z_l  - z_q)
\Big]~.
\label{zz}
\eeq
Note that those $C^X_i$ absent in the above have been fixed by anomaly cancellation conditions.\footnote{This fixes six constants: $C^d_2=C^e_2=-C^L_2$ and $C^L_1=3C^e_3=-3C^L_3=3(4C^d_3+C_1^e)$ and $C^d_1=-2C_1^e-9C^d_3$.} Finally, the multiplicities $N_\alpha$ (for $\alpha=1,2$) of the SM singlet RH-fermions $\nu_R^{(\alpha)}$ required by anomaly cancellation are given by
\beq
N_1&=
\frac{1}{3} \Big(3 n_G z_{l}^3-12 n_G z_{l}+12 n_G z_{q}^3-48 n_G z_{q}+3
   z_{d_L}^3-12 z_{d_L}+2z_{{\rm DM}}^3-8 z_{{\rm DM}}\\
   &~~-3 z_{d_R}^3+12
   z_{d_R}+z_{e_L}^3-4 z_{e_L}-z_{e_R}^3+4 z_{e_R}+2 z_{L_L}^3-8 z_{L_L}-2 z_{L_R}^3+8
   z_{L_R}\Big)\\[3pt]
N_2&=
\frac{1}{2} \Big(N_1+2 z_{\rm DM}+3 n_G z_{l}+12 n_G z_{q}+3 z_{d_L}-3
   z_{d_R}+z_{e_L}-z_{e_R}+2 z_{L_L}-2 z_{L_R}\Big)~.
\\
\eeq
Anomaly-free spectra can be found by choosing U(1)$'$ charges for the  SM fermions (provided $N_1,N_2\in\mathbb{Z}$), but are not unique and may not be the most minimal. 
Specifically, one obtains Model $\sharp$2b for
\beq
n_G=3,\qquad &z_{\rm DM}=z_l=z_q=1, \qquad
C_2^{L}=-2C_3^{d}=\frac{2}{3}~C_1^{e}~.
\eeq
The above parameter values leads to $N_1=-687$ and $N_2=-350$ but these can be manipulated to obtain the set of RH neutrinos in Table \ref{Tab7} using the method described in  \cite{Batra:2005rh}.


\end{document}